\documentclass[a4paper,12pt]{article}

\usepackage{jheppub}
\usepackage{etoolbox}
\usepackage{slashed}
\usepackage{subfig}
\usepackage{xcolor}
\usepackage{booktabs}
\usepackage{adjustbox}
\usepackage{dsfont}
\usepackage{ulem}
\usepackage{extarrows,cancel}
\usepackage{mathtools}
\usepackage{cases}
\usepackage{bm}       
\usepackage{bbm}
\usepackage{circledsteps}
\usepackage{hhline}

\usepackage{threeparttable}
\usepackage{multirow}
\usepackage{fnpct}

\usepackage{orcidlink}
\usepackage{lineno}

\DeclareFontFamily{OT1}{jheptitle}{}
\DeclareFontShape{OT1}{jheptitle}{bx}{n}{<->cmssbx10}{}
\makeatletter
\gdef\@fpheader{}
\patchcmd{\maketitle}%
  {\LARGE\flushleft\sffamily\bfseries}%
  {\flushleft\fontsize{19}{23}\usefont{OT1}{jheptitle}{bx}{n}}%
  {}{\errmessage{failed to patch the title size}}
\makeatother

\pgfkeys{/csteps/inner color=white}
\pgfkeys{/csteps/fill color=black}

\newcommand{\SM}{{\rm SM}}
\newcommand{\NP}{{\rm NP}}

\newcommand{\GeV}{{\,\rm GeV}}
\newcommand{\TeV}{{\,\rm TeV}}

\newcommand{\ab}{{\, \rm ab}}

\newcommand{\mB}{{\mathcal B}}
\newcommand{\mO}{{\mathcal O}}
\newcommand{\mC}{{\mathcal C}}
\newcommand{\mQ}{{\mathcal Q}}
\newcommand{\mL}{{\mathcal L}}
\newcommand{\mS}{{\mathcal S}}

\newcommand{\inv}{\text{inv}}
\newcommand{\DM}{\text{DM}}

\newcommand{\bv}{\boldsymbol{v}}
\newcommand{\bq}{\boldsymbol{q}}
\newcommand{\bp}{\boldsymbol{p}}
\newcommand{\vt}{{\bv_T^\perp}}
\newcommand{\vN}{{\bv_N^\perp}}

\newcommand{\SN}{\boldsymbol{S}_N}

\newcommand{\SDM}{\boldsymbol{S}_{\rm DM}}

\newcommand{\tgray}[1]{\textcolor{gray}{#1}}

\newcommand{\mpp}[1]{\textcolor{blue!50!black}{#1}}
\newcommand{\mpr}[1]{\textcolor{green!50!black}{#1}}

\newcommand{\hc}{\mathrm{h.c.}}
\newcommand{\nn}{\nonumber}

\newcommand{\boxtitle}[1]{\noindent\boxed{{#1}}}

\allowdisplaybreaks[4]

\title{Light top-flavoured dark matter in dark SMEFT: one-loop matching and phenomenology}

\author[a]{Meng-Chao Gao\orcidlink{0009-0000-5012-4907},}
\author[a]{Xin-Qiang Li\orcidlink{0000-0002-3962-3577},}
\author[a,b]{Ya-Dong Yang\orcidlink{0000-0002-9327-0557},}
\author[a]{Xing-Bo Yuan\orcidlink{0000-0002-5647-2563}}
\author[c]{and Xin Zhang\orcidlink{0000-0001-8051-1940}}

\affiliation[a]{Institute of Particle Physics and Key Laboratory of Quark and Lepton Physics~(MOE), Central China Normal University, Wuhan, Hubei 430079, China}
\affiliation[b]{Institute of Particle and Nuclear Physics, Henan Normal University, Xinxiang 453007, China}
\affiliation[c]{School of Physics, Hubei University, Wuhan, Hubei 430062, China}

\emailAdd{gmc8023@mails.ccnu.edu.cn}
\emailAdd{xqli@ccnu.edu.cn}
\emailAdd{yangyd@ccnu.edu.cn}
\emailAdd{y@ccnu.edu.cn}
\emailAdd{xinzhang@hubu.edu.cn}

\abstract{We study light top-flavoured dark matter (DM) below the electroweak (EW) scale within the dark Standard Model effective field theory (DSMEFT). We perform the one-loop matching of the DSMEFT onto the dark low-energy effective field theory (DLEFT) at the EW scale for all relevant operators that couple scalar, fermionic or vector DM, as well as axion-like particles, exclusively to third-generation quarks. After QCD evolution to the relevant low-energy scales, we match the DLEFT onto the non-relativistic EFT for DM--nucleus scattering. This EFT chain provides a unified framework for phenomenological studies of rare $B$- and $K$-meson decays with missing energy, direct and indirect detection, and the thermal relic abundance. Interpreting the Belle~II excess in $B^+\to K^+\nu\bar\nu$ as a decay into DM, we find that only the DSMEFT operators contributing at tree level can accommodate it with $\mathcal{O}(1)$ couplings for a new-physics scale of $1\,\mathrm{TeV}$. The decay $B^0\to K^{*0}+\mathrm{inv}$ provides the strongest constraint over large portions of the Belle~II-favoured parameter space, while $K^+\to\pi^++\mathrm{inv}$ can dominate at low and $B_s\to\inv$, $B^+\to\pi^++\mathrm{inv}$ and $\Upsilon(1S)\to\mathrm{inv}$ become important at high DM masses, respectively. Direct-detection searches by CRESST-III and DarkSide-50 are competitive for operators involving scalar quark currents and, depending on the operator, exclude the Belle~II-favoured band for $m_{\mathrm{DM}}\gtrsim1$--$2\,\mathrm{GeV}$. We also perform detailed analyses of the thermal relic abundance and the indirect-detection constraints from the CMB and Fermi-LAT.}

\begin{document}

\date{}
\maketitle

\section{Introduction}

Astrophysical and cosmological observations have firmly established that about $85\%$ of the matter in the Universe consists of non-baryonic dark matter (DM), with a relic abundance of $\Omega_{\DM}h^2=0.1200\pm0.0012$~\cite{Planck:2018vyg}, but its particle nature remains unknown~\cite{Bertone:2004pz,Cirelli:2024ssz}. For decades, the leading paradigm has been the weakly interacting massive particle (WIMP), whose mass lies close to the EW scale and whose abundance is set by thermal freeze-out~\cite{Lee:1977ua,Srednicki:1988ce,Gondolo:1990dk,Jungman:1995df}. Direct-detection experiments such as XENONnT~\cite{XENON:2023cxc,XENON:2025vwd}, PandaX-4T~\cite{PandaX:2024qfu} and LZ~\cite{LZ:2022lsv,LZ:2024zvo,LZ:2025igz} now exclude large parts of the WIMP parameter space for DM masses from a few $\GeV$ up to several $\TeV$~\cite{Arcadi:2017kky}. These increasingly stringent constraints have shifted the focus towards lighter DM candidates. Below a few $\GeV$, however, the nuclear recoil energies become too small for these detectors, and the sensitivity of direct detection deteriorates rapidly. Considerable efforts have been made in this direction, including conducting searches with dedicated low-threshold experiments such as CRESST-III~\cite{CRESST:2019jnq} and DarkSide-50~\cite{DarkSide:2018bpj}, and exploring DM--electron scattering as a potential direct-detection channel for DM coupled to electrons~\cite{Essig:2011nj,Graham:2012su,Essig:2015cda}.

DM with a mass of $\mO(\text{GeV})$ can also be produced in rare decays of $B$ and $K$ mesons, where it escapes detection and appears as missing energy~\cite{Bird:2004ts,Badin:2010uh,Kamenik:2011vy,He:2022ljo,BaBar:2024qqx}. In the SM, the corresponding flavour-changing neutral-current (FCNC) decays $M_1\to M_2\nu\bar\nu$ are loop and GIM suppressed and theoretically clean~\cite{Buchalla:1993bv,Buras:2014fpa,Brod:2021hsj,Becirevic:2023aov}. Together with the growing data sets at Belle~II~\cite{Belle-II:2022cgf} and NA62~\cite{NA62:2017rwk}, these features allow even small contributions from light invisible particles to be probed. In particular, these decays are sensitive to DM masses up to a few $\GeV$, precisely where direct detection loses sensitivity. Interest in this possibility has been revived by the first evidence for $B^+\to K^+\nu\bar\nu$ at Belle~II, with a branching ratio of $(23\pm7)\times10^{-6}$~\cite{Belle-II:2023esi}, about $2.7\sigma$ above the SM prediction~\cite{Buras:2014fpa,Becirevic:2023aov}. Besides heavy new physics (NP) modifying the $b\to s\nu\bar\nu$ transition~\cite{Allwicher:2023xba,Bause:2023mfe,He:2023bnk,Hou:2024vyw}, this excess can be explained by an additional decay channel into light invisible particles such as DM~\cite{McKeen:2023uzo,Altmannshofer:2023hkn,Datta:2023iln,Abdughani:2023dlr,Berezhnoy:2023rxx,Wang:2023trd,He:2023bnk,He:2024iju,Ho:2024cwk,Hou:2024vyw,Mescia:2026xju,Hong:2026qoj,Kim:2025zaf,Calibbi:2025rpx}. Such explanations are in turn constrained by the upper bounds on the other $B\to K^{(*)}\nu\bar\nu$ and $B\to\pi(\rho)\nu\bar\nu$ modes~\cite{Belle:2017oht}, by the measurements of and bounds on $K\to\pi\nu\bar\nu$~\cite{Chang:2026vvx,KOTO:2024zbl}, and by the limits on invisible $B_{s,d}$ decays~\cite{Alonso-Alvarez:2023mgc}. They will be further tested at Belle~II with $5\ab^{-1}$~\cite{Belle-II:2018jsg,Belle-II:2022cgf} and at future Tera-$Z$ factories~\cite{Li:2022tov,Amhis:2023mpj}.

Two complementary approaches are commonly used to describe light DM: simplified models, which introduce explicit mediators, and effective field theories (EFTs), in which heavy mediators are integrated out and their effects are organized in an expansion of higher-dimensional operators. We adopt the latter framework, with the DM fields retained as dynamical degrees of freedom down to the energies probed by experiments. Above the EW scale, this is the dark SMEFT (DSMEFT)~\cite{Cao:2009uw,Aebischer:2022wnl,Song:2023jqm,Liang:2023yta}, which extends the SMEFT~\cite{Buchmuller:1985jz,Grzadkowski:2010es} by SM gauge-singlet DM fields of spin $0$, $1/2$ and $1$, and which can be supplemented by axion-like particles (ALPs)~\cite{Georgi:1986df,Brivio:2017ije,Bauer:2020jbp,Song:2023lxf}. Below the EW scale, the top quark, the Higgs boson, and the $W$ and $Z$ bosons are all integrated out. The interactions of the light DM with quarks, gluons and photons are then described by the dark LEFT (DLEFT)~\cite{Aebischer:2022wnl}, the corresponding extension of the LEFT~\cite{Jenkins:2017jig,Jenkins:2017dyc}. At the hadronic scale, the DM--nucleus scattering is described by a non-relativistic EFT (NREFT) of DM--nucleon interactions~\cite{Fan:2010gt,Fitzpatrick:2012ix,Anand:2013yka}. Relating a DSMEFT operator to rare meson decays, DM--nucleus scattering, DM annihilation in the early Universe and the resulting relic density therefore requires a chain of EFTs: the matching of the DSMEFT onto the DLEFT at the EW scale, the renormalization-group (RG) evolution within the DLEFT, and the matching of the DLEFT onto the NREFT. While the tree-level matching between the DSMEFT and DLEFT has been derived in ref.~\cite{Aebischer:2022wnl}, this work focuses on the one-loop matching, particularly on the contributions generated by integrating out the top quark.

Importantly, the complementarity between rare meson decays and direct detection concerns not only their different DM-mass reach but also the flavour structure of the interactions they probe. Rare meson decays with missing energy probe flavour-changing quark currents coupled to DM, whereas DM--nucleus scattering is controlled primarily by flavour-diagonal interactions of DM with light quarks and gluons. In a concrete UV completion of the dark sector, however, these couplings need not be independent and are often correlated. It is therefore well motivated to consider specific flavour structures for the interactions between DM and SM fermions within the EFT. Indirect detection and the relic density provide complementary information through DM annihilation, with the relic density potentially receiving contributions from both flavour-changing and flavour-conserving interactions. A combined analysis of meson decays, direct and indirect detection, and the relic density can thus probe the flavour structure of DM interactions with SM fermions.

In this context, a particularly well-motivated scenario is the top-flavoured (or top-philic) DM, whose interactions with SM fermions are restricted to the third-generation quarks. This scenario has been extensively studied in the literature~\cite{Cheung:2010zf,Agrawal:2011ze,Kumar:2013hfa,Gomez:2014lva,Kilic:2015vka,Haisch:2015ioa,Blanke:2017tnb,Garny:2018icg,Cornell:2021crh,Liu:2021crr,Hu:2024xes}. Most studies, however, focus on DM masses at or above the EW scale, where collider searches and direct detection provide the main probes. For light DM, the situation is qualitatively different. Since the top quark is absent from the low-energy theory while the DM is not, all low-energy effects arise from matching at the EW scale. At low energies, loops involving top quarks and $W$ bosons can induce FCNC interactions between DM and down-type quarks, thereby mediating rare $B$- and $K$-meson decays. Top-quark loops also generate couplings to gluons and photons and, for most operators, flavour-diagonal couplings to light quarks. Together, these interactions control direct detection, DM annihilation and indirect detection. Light top-flavoured DM thus provides a minimal and predictive scenario, in which flavour, direct-detection and cosmological observables are correlated through a single coupling per operator, and in which the one-loop DSMEFT--DLEFT matching determines most of the phenomenology.


In this work, we study the top-flavoured scenario for light DM with masses within the kinematic reach of rare $B$-meson decays, in the DSMEFT framework. Our analysis covers scalar, fermionic and vector DM, as well as ALPs. In particular, we perform the one-loop matching of the DSMEFT onto the DLEFT at the EW scale. The resulting DLEFT Wilson coefficients are evolved to the relevant low-energy scales taking into account the dominant QCD effect, and the DLEFT is subsequently matched onto the NREFT for DM--nucleus scattering. This EFT chain is then used for a comprehensive phenomenological analysis of rare $B$- and $K$-meson decays, direct-detection constraints, the relic abundance, and CMB~\cite{Boddy:2015efa} and Fermi-LAT~\cite{Ackermann:2015lka} constraints on DM annihilation. This extends our previous analysis in ref.~\cite{Hou:2024vyw}, in which the DSMEFT operators with generic or minimally flavour-violating (MFV)~\cite{DAmbrosio:2002vsn} structures were matched onto the DLEFT at tree level.

The remainder of this paper is organized as follows. In section~\ref{sec:Theoretical framework}, we introduce the DSMEFT and DLEFT operators considered in this work, discuss the flavour structure of the top-flavoured scenario, and present the QCD RG evolution of the DLEFT Wilson coefficients. The matching between the DSMEFT and the DLEFT is described in section~\ref{sec:matching}, with the explicit matching coefficients collected in appendix~\ref{app:matching coefficients}. Sections~\ref{sec:flavour}--\ref{sec:indirect detection} are devoted to the individual observables: rare $B$- and $K$-meson decays in section~\ref{sec:flavour}; direct detection, including the DLEFT--NREFT matching, in section~\ref{sec:direct detection}; the relic density in section~\ref{sec:Relic Density}; and indirect detection in section~\ref{sec:indirect detection}. The combined numerical analysis is presented in section~\ref{sec:Numerical analysis}, and we conclude in section~\ref{sec:conclusion}. The projected flavour constraints for all representative operators are shown in appendix~\ref{app:projections}. For convenience, we collect in appendix~\ref{app:hadronic matrix element} the hadronic matrix elements for rare $B$- and $K$-meson decays, as well as for direct detection.

\section{Theoretical framework}
\label{sec:Theoretical framework}

In this section, we first recapitulate the DSMEFT and DLEFT frameworks, and then derive the EW-scale matching of the Wilson coefficients from the DSMEFT onto the DLEFT for the top-loop induced processes $M_1\to M_2+\DM\,(+\DM)$.

\subsection{DSMEFT}
\label{sec:DSMEFT}

The DSMEFT extends the SMEFT~\cite{Buchmuller:1985jz,Grzadkowski:2010es} to incorporate the SM gauge-singlet DM particles with masses either above or below the EW scale~\cite{Cao:2009uw,Aebischer:2022wnl,Song:2023jqm}. Following the conventions of ref.~\cite{Aebischer:2022wnl}, the DSMEFT Lagrangian reads
\begin{align}\label{eq:DSMEFT:Lagrangian}
    \mL_{\mathrm{DSMEFT}}\supset \sum_i\mC_i\mQ_i^{(4)}+\frac{1}{\Lambda}\sum_{j}\mC_j \mQ_{j}^{(5)} + \frac{1}{\Lambda^2}\sum_{k} \mC_k \mQ_{k}^{(6)}+ \frac{1}{\Lambda^3}\sum_{l} \mC_l \mQ_{l}^{(7)},
\end{align}
where $\Lambda$ is the NP scale above which new degrees of freedom appear and are coupled to both the SM and DM sectors. The dimention-$d$ operators $\mQ_i^{(d)}$ involve at least one DM field and respect the SM gauge symmetry, while the dimensionless Wilson coefficients $\mC_i$ are defined at the EW scale $\mu_{\rm EW}$. All operators are written in the flavour eigenbasis.

In the DSMEFT framework, the DM field can be a spin-0 scalar $\phi$, a spin-1/2 fermion $\chi$, or a spin-1 vector $X$, and each type may carry multiple generations. Throughout this work, we assume a single DM generation; for operators that vanish in this limit, two generations are considered instead. In the top-flavoured DM scenario, the DM contributes to rare $B$- and $K$-meson decays via top-quark loops, so only operators involving up-type quarks need to be included.\footnote{In general, above the EW scale, the RG running also mixes the top-quark operators into the ones that contain the Higgs or gauge bosons rather than the top quarks~\cite{Chala:2020pbn,Datta:2020ocb,Han:2020pff}. Since the DSMEFT is fixed at the EW scale in this work, such operators do not enter our analysis. \label{fn:DSMEFTatmuEW}} Motivated by the large excess of $\mathcal B(B^+\to K^+\nu\bar \nu)$ reported by Belle II~\cite{Belle-II:2023esi}, we restrict ourselves to operators of dimension $\leq 6$, which are less suppressed than higher-dimensional ones, together with the single dimension-7 operator that generates a Yukawa-type interaction for fermionic DM. In addition, we include the ALPs~\cite{Peccei:1977hh,Peccei:1977ur,Weinberg:1977ma,Wilczek:1977pj} within the same EFT framework~\cite{Georgi:1986df,Brivio:2017ije,Chala:2020wvs,Bauer:2020jbp,Galda:2021hbr,Song:2023lxf}; ALPs are a special class of pseudo Nambu-Goldstone bosons that can arise from spontaneous breaking of a global symmetry.

\begin{table}[t]
    \renewcommand*{\arraystretch}{1.3}
	\tabcolsep 0.21in
    \centering
    \begin{tabular}{c | c | c | c}
    \toprule
    operator & definition & operator & definition 
    \\\midrule
    \multirow{2}{*}{$\mQ_{u\phi}$} & \multirow{2}{*}{$(\bar q_p u_r\widetilde H)\phi+\hc$} & $\mQ_{qX}$ & $(\bar{q}_p\gamma^{\mu}q_r)X_{\mu}$
    \\
    & & $\mQ_{uX}$ & $(\bar{u}_p\gamma^{\mu}u_r)X_{\mu}$
    \\
    \multirow{2}{*}{$\mQ_{u\phi^2}$} & \multirow{2}{*}{$(\bar q_p u_r\widetilde H)\phi^2 +\hc$} & $\mQ_{HuX}$ & $(H^{\dagger}H)(\bar{u}_p\gamma^{\mu}u_r)X_{\mu}$
    \\
    & & $\mQ_{HqX}^{(1)}$ & $(H^{\dagger}H)(\bar{q}_p\gamma^{\mu}q_r)X_{\mu}$
    \\
    \multirow{2}{*}{$\mQ_{\phi q}$} & \multirow{2}{*}{$(\bar q_p\gamma_\mu q_r)(i\phi_1\overleftrightarrow{\partial^\mu}\phi_2)$} & $\mQ_{HqX}^{(3)}$ & $(H^{\dagger}\tau^I H)(\bar{q}_p\tau^I \gamma^{\mu}q_r)X_{\mu}$
    \\
    & & $\mQ_{uX^2}$ & $(\bar q_p u_r\widetilde H)  X_\mu X^\mu+\hc$
    \\
    \multirow{2}{*}{$\mQ_{\phi u}$} & \multirow{2}{*}{$(\bar u_p\gamma_\mu u_r)(i\phi_1\overleftrightarrow{\partial^\mu}\phi_2)$} & $\mQ_{qXX}$ & $(\bar q_p\gamma_\mu q_r)X^{\mu\nu}X_\nu$
    \\
    & & $\mQ_{q\widetilde X X}$ & $(\bar q_p\gamma_\mu q_r)\widetilde X^{\mu\nu}X_\nu$
    \\ \hhline{--|~|~}
    \multirow{2}{*}{$\mQ_{q\chi}$} & \multirow{2}{*}{$(\bar q_p\gamma_\mu q_r)(\bar \chi\gamma^\mu\chi)$} & $\mQ_{uXX}$ & $(\bar u_p\gamma_\mu u_r)X^{\mu\nu}X_\nu$
    \\
    & & $\mQ_{u\widetilde X X}$ & $(\bar u_p\gamma_\mu u_r)\widetilde X^{\mu\nu}X_\nu$
    \\
    \multirow{2}{*}{$\mQ_{u\chi}$} & \multirow{2}{*}{$(\bar u_p\gamma_\mu u_r)(\bar \chi\gamma^\mu\chi)$} & $\mQ_{D qX^2}$ & $\left(\bar q_p\gamma^\nu i \overleftrightarrow{D^\mu} q_r \right) X_\mu X_\nu$
    \\
    & & $\mQ_{D uX^2}$ & $\left(\bar u_p\gamma^\nu i \overleftrightarrow{D^\mu} u_r \right) X_\mu X_\nu$
    \\
    \multirow{2}{*}{$\mQ_{u\chi^2}$} & \multirow{2}{*}{$(\bar q_p u_r \widetilde H)(\chi^T C \chi)$} & $\mQ_{uHX^2}$ & $(\bar q_p \sigma_{\mu\nu} u_r\widetilde H)  X^\mu_1 X^\nu_2+\hc$
    \\
    & & $\mQ_{uHX}$ & $(\bar{q}_p\sigma_{\mu\nu}u_r) \widetilde{H}X^{\mu\nu}+\hc$
    \\ \hhline{--|~|~}
    $\mQ_{qa}$ & $(\bar q_p \gamma_{\mu} q_r) \partial^\mu a$ & &
    \\
    $\mQ_{ua}$ & $(\bar u_p \gamma_{\mu} u_r) \partial^\mu a$ & &
    \\\bottomrule
    \end{tabular}
    \caption{DSMEFT operators considered in this work, given in the flavour eigenbasis. The four blocks correspond to the cases with the DM field being a scalar ($\phi$), a fermion ($\chi$), an ALP ($a$), or a vector ($X^\mu$), respectively. See text for more details.}
    \label{tab:DSMEFT}
\end{table}

All DSMEFT operators considered in this work are collected in table~\ref{tab:DSMEFT}. The fields $q$ and $u$ stand for the left-handed quark doublet and the right-handed up-type quark singlet in the flavour eigenbasis, respectively; $p,r$ are generation indices. $H$ is the SM Higgs doublet and $\widetilde H\equiv i\tau^2 H^*$. The DM fields are represented by the real scalar $\phi$ (or $\phi_i$, $i=1,2$, in the two-generation case), the fermion $\chi$, the real vector $X^\mu$, and the ALP $a$. Without loss of generality, $\chi$ is taken to be a right-handed Dirac field, since the observables studied throughout this work are insensitive to its chirality. For the vector DM, we follow ref.~\cite{Song:2023jqm} and take both $X^\mu$ and its field strength tensor $X_{\mu\nu}\equiv \partial_\mu X_\nu-\partial_\nu X_\mu$ (with the dual tensor given by $\widetilde X_{\mu\nu}\equiv \tfrac{1}{2}\epsilon_{\mu\nu\rho\sigma}X^{\rho\sigma}$) as the building blocks.\footnote{To make the operator basis complete in flavour space, we follow the approach proposed in refs.~\cite{Li:2020gnx,Degrande:2023iob} and obtain $\mQ_{DuX^2}$ and $\mQ_{DqX^2}$ by proper Hermitianization of the operators in ref.~\cite{Song:2023jqm}.} Among these operators, $\mQ_{\phi q}$, $\mQ_{\phi u}$, and $\mQ_{uHX^2}$ vanish in the single-generation limit and therefore require at least two DM generations. The remaining conventions are $\phi_1\overleftrightarrow{\partial^\mu}\phi_2\equiv \phi_1(\partial^\mu\phi_2)-(\partial^\mu\phi_1)\phi_2$, $\sigma_{\mu\nu}\equiv \tfrac{i}{2}[\gamma_\mu,\gamma_\nu]$, $C\equiv i\gamma^2\gamma^0$ for the charge-conjugation matrix, and $D_\mu\equiv \partial_\mu+ig_s T^a G_\mu^a+ig_1 Y B_\mu$ for the $SU(3)_C\otimes U(1)_Y$ covariant derivative, with $T^a=\lambda^a/2$ and $\lambda^a$ the Gell-Mann matrices. For the non-Hermitian operators in table~\ref{tab:DSMEFT}, the abbreviation ``h.c.'' indicates that the Hermitian conjugate is to be added to the Lagrangian in eq.~\eqref{eq:DSMEFT:Lagrangian}.

The decay rates of rare $B$- and $K$-meson processes induced by operators involving the vector DM field $X^\mu$ diverge in the massless limit $m_X\to 0$. This divergence originates from the longitudinal polarization of $X^\mu$ and is tied to the well-known fact that defining a massless limit for a vector field without an underlying gauge symmetry is problematic~\cite{Kamenik:2011vy}. Following refs.~\cite{Kamenik:2011vy,He:2022ljo,Williams:2011qb}, we assume some kind of  Higgs mechanism in the dark sector. Consequently, the Wilson coefficients of the vector-DM operators in table~\ref{tab:DSMEFT} are proportional to powers of $m_X/\Lambda$ after spontaneous symmetry breaking, and can be rescaled as
\begin{align} \label{eq:rescaledWCs}
    \mC_i=\tilde\mC_i \cdot
    \begin{cases}
    (m_X/\Lambda)^2
    &
    \text{for}~\mQ_i=\mQ_{uX^2},\mQ_{DuX^2},\mQ_{DqX^2},\mQ_{uHX^2},
    \\
    1
    &
    \text{for}~\mQ_i=\mQ_{uHX},
    \\
    (m_X/\Lambda)
    &
    \text{for}~\mQ_i=\text{others}.
    \end{cases}
\end{align}
In what follows, we work with these rescaled Wilson coefficients $\tilde\mC_i$.

In the DSMEFT, the quark currents in the flavour eigenbasis can be divided into the following four categories:
\begin{align}
    \bar q \gamma^\mu \mC q,
    \qquad
    \bar u \gamma^\mu \mC u,
    \qquad
    \bar q \mC u,
    \qquad
    \bar q \sigma^{\mu\nu}\mC u,
\end{align}
where $\mC$ is the Wilson coefficient of the operator $\mQ$ and is represented by a $3\times 3$ matrix in quark flavour space.\footnote{Here and in the following, we suppress the operator index $i$, so that $\mC$ represents $\mC_i$.} 
Specific to the top-flavoured DM scenario, the Wilson coefficient takes the form
\begin{align}
    \mC_t \equiv 
    \begin{pmatrix}
        0 & 0 & 0  \\
        0 & 0 & 0  \\
        \;0\; & \;0\; &\mC_{33}
    \end{pmatrix}.
\end{align}
In the following, we rotate these four currents to the mass eigenbasis and derive explicit expressions of their Wilson coefficients in the top-flavoured DM scenario. As in ref.~\cite{Bissmann:2020mfi}, without loss of generality, we choose to work in the so-called ``up basis"~\cite{Brivio:2020onw,Sun:2025axx,Datta:2025csr}. In this basis, the flavour and mass eigenstates of $u_R$ and $d_R$ are identical, while $q$ takes the form
\begin{align}\label{eq:up basis}
    q=
        \begin{pmatrix}
        u_L\\
        V d_L
        \end{pmatrix},
\end{align}
where $V$ denotes the CKM matrix, and $u_L$ and $d_L$ are the left-handed up and down-type quarks in the mass eigenbasis, respectively.

By using eq~\eqref{eq:up basis}, the current $ \bar q \gamma^{\mu} \mC q$ can be written in the mass eigenbasis as
\begin{align}
    \bar q \gamma^\mu \mC_t q
    =  \bar u_L \gamma^\mu \mC_t u_L
      +\bar d_L \gamma^\mu V^\dagger \mC_t V d_L.
\end{align}
Similarly, one can obtain the basis rotations for the currents $\bar q \mC u$ and $\bar q \sigma^{\mu\nu}\mC u$ as
\begin{align}
    \bar q \mC_t u=&\left(\bar u_L \mC_t u_R,\; \bar d_L V^{\dagger}\mC_t u_R\right),\\
    \bar q \sigma^{\mu\nu}\mC_t u=&\left(\bar u_L\sigma^{\mu\nu}\mC_t u_R,\; \bar d_L \sigma^{\mu\nu} V^{\dagger}\mC_t u_R\right).
\end{align}
For the right-handed quark current $\bar u \gamma^\mu \mC u$, no basis rotation is needed, and we have
\begin{align}
    \bar u \gamma^\mu \mC_t u = \bar u_R \gamma^\mu \mC_t u_R.
\end{align}
Finally, in the top-flavoured DM scenario, the Wilson coefficients for the various quark currents in the mass eigenbasis are summarized as
\begin{align}\label{eq:WC:quark current}
    \mC=
    \begin{cases}
        \mC_t   &\text{for}~ \bar u_L \gamma^\mu \mC u_L,\, u_R \gamma^\mu \mC u_R,\, \bar u_L \mC u_R,\, \bar u_L \sigma^{\mu\nu}\mC u_R,\\
        V^\dagger \mC_t V  &\text{for}~ \bar d_L \gamma^\mu \mC d_L,\\
        V^\dagger \mC_t & \text{for}~ \bar d_L \mC u_R,\, \bar d_L \sigma^{\mu\nu} \mC u_R.  
    \end{cases}
\end{align}
It is clear that the FCNC couplings between the left-handed down-type quarks origin from the $SU(2)_L$ gauge invariance and are controlled by the matrix $V^\dagger \mC_t V$. Its explicit expression and numerical values are given by
\begin{align}
    V^{\dagger}\mC_t V = &
    \mC_{33}
    \begin{pmatrix}
        V_{td}^* V_{td}&V_{td}^* V_{ts}&V_{td}^* V_{tb} \\
        V_{ts}^* V_{td}&V_{ts}^* V_{ts}&V_{ts}^* V_{tb} \\
        V_{tb}^* V_{td}&V_{tb}^* V_{ts}&V_{tb}^* V_{tb}
    \end{pmatrix}\nn\\
    = &
    \mC_{33}
    \begin{pmatrix}
        0.71 & \; -3.10-1.46i \; & 77.09+34.43i \\
        -3.10+1.46i & \; 16.43 \; & -404.86+7.90i \\
         77.09-34.43i & \; -404.86-7.90i \; & 9982.86
    \end{pmatrix}
    \times 10^{-4}.
\end{align}
Similarly, the charged currents are controlled by $V^\dagger \mC_t$, whose expression and numerical values are given by
\begin{align}
    V^{\dagger}\mC_t=
    \mC_{33}
    \begin{pmatrix}
    \;0\; & \;0\; & V_{td}^*  \\
    0 & 0 & V_{ts}^*  \\
    0 & 0 & V_{tb}^*
    \end{pmatrix}  
    =
    \mC_{33}
    \begin{pmatrix}
        0 & 0 & 0.77+0.34i \\
        0 & 0 & -4.05+0.08i \\
        0 & 0 & 99.91
    \end{pmatrix}
    \times 10^{-2}.
\end{align}

We can see from the above discussions that, in the top-flavoured DM scenario, the Wilson coefficients of the DSMEFT operators should take special flavour structures in the mass eigenbasis. Explicitly, based on eq.~\eqref{eq:WC:quark current}, we obtain
\begin{align}\label{eq:WC:DSMEFT operator}
    \mC_i=
    \begin{cases}
        \mC_t^i  
        &
        \text{for}~ \mQ_i=\mQ_{\phi u},\,\mQ_{u\chi},\, \mQ_{uX},\, \mQ_{HuX},\, \mQ_{uXX},\, \mQ_{u\widetilde XX},\, \mQ_{DuX^2},\, \mQ_{ua},
        \\
        (\mC_t^i,\,V^\dagger \mC_t^i)  
        &
        \text{for}~ \mQ_i=\mQ_{u\phi},\, \mQ_{u\phi^2},\, \mQ_{u\chi^2},\, \mQ_{uHX},\, \mQ_{uX^2},\, \mQ_{uHX^2},
        \\
        (\mC_t^i,\,V^\dagger \mC_t^i V)  
        &
        \text{for}~ \mQ_i=\mQ_{\phi q},\, \mQ_{q\chi},\, \mQ_{qX},\, \mQ_{HqX}^{(1,3)},\, \mQ_{qXX},\, \mQ_{q\widetilde XX},\, \mQ_{DqX^2},\,\mQ_{qa},
    \end{cases}
\end{align}
where the first and second terms in the parenthesis denote the flavour structures for the quark currents involving $u_L$ and $d_L$, respectively.

\subsection{DLEFT}
\label{subsec:DLEFT}

Below the EW scale, the interactions between the light SM and the light DM particles can be described by the DLEFT~\cite{Aebischer:2022wnl,Liang:2023yta}. This EFT is constructed by integrating out all the heavy particles with masses at or above the EW scale and can be considered as a
generalization of the LEFT~\cite{Buchalla:1995vs,Jenkins:2017jig} by including the light DM particles. The DLEFT respects the $SU(3)_C \otimes U(1)_{\text{em}}$ gauge symmetry of the SM after spontaneous symmetry breaking. Following the convention of ref.~\cite{Aebischer:2022wnl}, the DLEFT Lagrangian takes the form
\begin{align}
    \mL_{\text{DLEFT}}\supset \sum_{i}L_i \mO_{i}^{(4)}+\frac{1}{v}\sum_{j}L_j \mO_{j}^{(5)} + \frac{1}{v^2}\sum_{k} L_k \mO_{k}^{(6)}+ \frac{1}{v^3}\sum_{l} L_l \mO_{l}^{(7)}, 
\end{align}
where $v=246\GeV$ denotes the vacuum expectation value. $\mathcal O_i^{(d)}$ are the dimension-$d$ effective operators and $L_i$ are the corresponding Wilson coefficients. Here, all the effective operators are defined in the mass eigenbasis. For the processes involved in sections~\ref{sec:flavour}--\ref{sec:indirect detection}, we list all the relevant DLEFT operators in table~\ref{tab:DLEFT}. For the operators involving vector DM, we have chosen the basis used in ref.~\cite{Hou:2024vyw}. In this basis, the combinations of $X^{\mu\nu}$ and $X^\mu$ are similar as in the DSMEFT operators of table~\ref{tab:DSMEFT}. One can prove that this basis is equivalent to the one adopted in ref.~\cite{He:2022ljo}.

\begin{table}[htbp]
    \renewcommand*{\arraystretch}{1.3}
	\tabcolsep 0.33in
    \centering
    \begin{adjustbox}{totalheight=0.88\textheight}
    \begin{tabular}{c | c | c | c | c}
    \toprule
    operator & definition & tree & loop & eq.
    \\\midrule
    $\mO_{d\phi}$ & $(\bar{d}_{Lp} d_{Rr})\phi+\hc$ &  $-$  &  $\mQ_{u\phi}$ & \eqref{eq:WC:S:1}
    \\
    $\mO_{d\phi^2}$ & $(\bar{d}_{Lp} d_{Rr})\phi^2+\hc$ & $-$ & \mpr{$\mQ_{u \phi^2}$}, \mpp{$\mQ_{u\phi^2}$} & \eqref{eq:WC:dphi2:pr}, \eqref{eq:WC:S:2:pp} 
    \\
    $\mO_{u\phi^2}$ & $(\bar{u}_{Lp} u_{Rr})\phi^2+\hc$ & $-$ & \mpp{$\mQ_{u\phi^2}$} & \eqref{eq:WC:uphi2:pp}
    \\
    $\mO_{\phi d}^{L}$ & $(\bar{d}_{Lp}\gamma_{\mu} d_{Lr})(i\phi_1\overleftrightarrow{\partial^\mu}\phi_2)$ & $\mQ_{\phi q}$ & \mpr{$\mQ_{\phi u}$} & \eqref{eq:WC:S2:pr}
    \\
    $\mO_{\phi d}^{R}$ & $(\bar{d}_{Rp}\gamma_{\mu} d_{Rr})(i\phi_1\overleftrightarrow{\partial^\mu}\phi_2)$ & $-$ & $-$ & \eqref{eq:WC:phid:R:pr} 
    \\\midrule
    $\mO_{d\chi}^{S,LR}$ & $(\bar{d}_{Rp}d_{Lr})(\chi^TC\chi)+\hc$ & $-$ & \mpr{$\mQ_{u\chi^2}$}, \mpp{$\mQ_{u\chi^2}$} & \eqref{eq:WC:dchi:SLR:pr}, \eqref{eq:WC:dchi:SLR:pp}
    \\
    $\mO_{u\chi}^{S,LR}$ & $(\bar{u}_{Rp}u_{Lr})(\chi^TC\chi)+\hc$ & $-$ & \mpp{$\mQ_{u\chi^2}$} & \eqref{eq:WC:uchi:SLR:pp}
    \\
    $\mO_{d\chi}^{S,RR}$ & $(\bar{d}_{Lp} d_{Rr})(\chi^TC\chi)+\hc$ & $-$ & \mpr{$\mQ_{u\chi^2}$}, \mpp{$\mQ_{u\chi^2}$} & \eqref{eq:WC:dchi:SRR:pr},  \eqref{eq:WC:dchi:SRR:pp}
    \\
    $\mO_{u\chi}^{S,RR}$ & $(\bar{u}_{Lp} u_{Rr})(\chi^TC\chi)+\hc$ & $-$ & \mpp{$\mQ_{u\chi^2}$} & \eqref{eq:WC:uchi:SRR:pp}
    \\
    $\mO_{d\chi}^{V,LR}$ & $(\bar{d}_{Lp}\gamma_{\mu} d_{Lr})(\bar{\chi}\gamma^{\mu}\chi)$ & $\mQ_{q\chi}$ & \mpr{$\mQ_{u\chi}$}, \mpp{$\mQ_{u\chi}$} & \eqref{eq:WC:dchi:VLR:pr}, \eqref{eq:WC:dchi:VLR:pp}
    \\
    $\mO_{u\chi}^{V,LR}$ & $(\bar{u}_{Lp}\gamma_{\mu} u_{Lr})(\bar{\chi}\gamma^{\mu}\chi)$ & $-$ & \mpp{$\mQ_{u\chi}$}, \mpp{$\mQ_{q\chi}$} & \eqref{eq:WC:uchi:VLR:pp}
    \\
    $\mO_{d\chi}^{V,RR}$ & $(\bar{d}_{Rp}\gamma_{\mu} d_{Rr})(\bar{\chi}\gamma^{\mu}\chi)$ & $-$ & \mpp{$\mQ_{u\chi}$} & \eqref{eq:WC:dchi:VRR:pp}
    \\
    $\mO_{u\chi}^{V,RR}$ & $(\bar{u}_{Rp}\gamma_{\mu} u_{Rr})(\bar{\chi}\gamma^{\mu}\chi)$ & $-$ & \mpp{$\mQ_{u\chi}$, $\mQ_{q\chi}$}, & \eqref{eq:WC:uchi:VRR:pp}
    \\
    $\mO_{d\chi}^{T,RR}$ & $(\bar{d}_{Lp}\sigma^{\mu\nu} d_{Rr})(\chi^TC\sigma_{\mu\nu}\chi)+\hc$ & $-$ & $-$ & \eqref{eq:WC:F:pr} 
    \\\midrule
    $\mO_{dX}^{T}$ & $(\bar{d}_{Lp}\sigma^{\mu\nu} d_{Rr})X_{\mu\nu}+\hc$ & $-$ & \mpr{$\mQ_{uX},\mQ_{qX},\mQ_{uHX},\mQ_{HuX},\mQ_{HqX}$} & \eqref{eq:WC:dX:T:pr}
    \\
    $\mO_{dX}^{L}$ & $(\bar{d}_{Lp}\gamma^{\mu} d_{Lr})X_{\mu}$ & $\mQ_{qX},\mQ_{HqX}$ & \mpr{$\mQ_{uX}, \mQ_{HuX}$} &  \eqref{eq:WC:dX:L:pr}
    \\
    $\mO_{dX}^{R}$ & $(\bar{d}_{Rp}\gamma^{\mu} d_{Rr})X_{\mu}$ & $-$ & $-$ &  \eqref{eq:WC:dX:R:pr}
    \\
    $\mO_{dX^2}$ & $(\bar{d}_{Lp} d_{Rr})X_{\mu}X^{\mu}+\hc$ & $-$ & \mpr{$\mQ_{uX^2},\mQ_{DuX^2},\mQ_{DqX^2}$}, \mpp{$\mQ_{uX^2}$} & \eqref{eq:WC:dX2:pr}, \eqref{eq:WC:dX2:pp}
    \\
    $\mO_{d X^2}^T$ & $(\bar d_{Lp} \sigma_{\mu\nu} d_{Rr})X^{\mu}_1X^{\nu}_2+\hc$ & $-$ & \mpr{$\mQ_{uHX^2}$}  & \eqref{eq:WC:dX2:T:pr}
    \\
    $\mO_{dXX}^{L}$ & $(\bar{d}_{Lp}\gamma^{\mu}  d_{Lr})X_{\mu\nu}X^{\nu}$ & $\mQ_{qXX}$ & \mpr{$\mQ_{uXX}$}, \mpp{$\mQ_{uXX}$}  & \eqref{eq:WC:dXX:L:pr}, \eqref{eq:WC:dXX:L:pp}
    \\
    $\mO_{dXX}^{R}$ & $(\bar{d}_{Rp}\gamma^{\mu} d_{Rr})X_{\mu\nu}X^{\nu}$ & $-$ & \mpp{$\mQ_{uXX}$}  & \eqref{eq:WC:dXX:R:pp}
    \\
    $\mO_{d\widetilde XX}^{L}$ & $(\bar{d}_{Lp}\gamma^{\mu} d_{Lr})\widetilde X_{\mu\nu}X^{\nu}$ & $\mQ_{q \widetilde X X}$ & \mpr{$\mQ_{u \widetilde X X}$} & \eqref{eq:WC:dXtX:L:pr}
    \\
    $\mO_{d\widetilde XX}^{R}$ & $(\bar{d}_{Rp}\gamma^{\mu} d_{Rr})\widetilde X_{\mu\nu}X^{\nu}$ & $-$ & $-$ & \eqref{eq:WC:dXtX:R:pr} 
    \\
    $\mO_{D dX^2}^{L}$ & $(\bar{d}_{Lp}\gamma^{\mu} i \overleftrightarrow D^{\nu} d_{Lr})X_{\mu}X_{\nu}$ & $\mQ_{DqX^2}$ &  \mpr{$\mQ_{DuX^2}$}, \mpp{$\mQ_{DuX^2}$}  & \eqref{eq:WC:DdX2:L:pr}, \eqref{eq:WC:DdX2:L:pp}
    \\
    $\mO_{D dX^2}^{R}$ & $(\bar{d}_{Rp}\gamma^{\mu} i \overleftrightarrow D^{\nu} d_{Rr})X_{\mu}X_{\nu}$ & $-$ & \mpp{$\mQ_{DuX^2}$}   & \eqref{eq:WC:DdX2:R:pp}
    \\\midrule
    $\mO_{da}^L$ & $(\bar d_{Lp} \gamma_{\mu} d_{Lr})\partial^{\mu} a$ & $\mQ_{qa}$ & \mpr{$\mQ_{ua}$} & \eqref{eq:WC:ALP:L:pr}
    \\
    $\mO_{da}^R$ & $(\bar d_{Rp} \gamma_{\mu} d_{Rr})\partial^{\mu} a$ & $-$ & $-$ & \eqref{eq:WC:ALP:R:pr}
    \\\midrule
    $\mO_{GG\phi^2}$ & $G_{\mu\nu}^a G^{\mu\nu,a}\phi^2$ & $-$ & $\mQ_{u\phi^2}$  & \eqref{eq:WC:S:glgl}
    \\
    $\mO_{G \widetilde G \phi^2}$ & $G_{\mu\nu}^a \widetilde G^{\mu\nu,a}\phi^2$ & $-$ & $\mQ_{u\phi^2}$  & \eqref{eq:WC:S:glgl}
    \\
    $\mO_{GG\chi}$ & $G_{\mu\nu}^a G^{\mu\nu,a}\chi^T C \chi$ & $-$ & $\mQ_{u\chi^2}$  & \eqref{eq:WC:F:glgl}
    \\
    $\mO_{G\widetilde{G}\chi}$ & $G_{\mu\nu}^a \widetilde G^{\mu\nu,a}\chi^T C \chi$ & $-$ & $\mQ_{u\chi^2}$  & \eqref{eq:WC:F:glgl}
    \\
    $\mO_{GGXX}$ & $G_{\mu\nu}^a G^{\mu\nu,a}X_{\rho}X^{\rho}$ & $-$ & $\mQ_{uX^2},\tgray{\mQ_{DqX^2},\mQ_{DuX^2}}$ & \eqref{eq:WC:V:glgl}
    \\
    $\mO_{G\widetilde{G}XX}$ & $G_{\mu\nu}^a \widetilde G^{\mu\nu,a}X_{\rho}X^{\rho}$ & $-$ & $\mQ_{uX^2},\tgray{\mQ_{DqX^2},\mQ_{DuX^2}}$ & \eqref{eq:WC:V:glgl}
    \\
    $\mO_{GXGX}$ & $G_{\lambda\mu}^a G^{\lambda,a}_{\;\;\nu}X^{\mu}X^{\nu}$ & $-$ & \tgray{$\mQ_{DqX^2},\mQ_{DuX^2}$} & \eqref{eq:WC:V:glgl}
    \\
    $\mO_{GX\widetilde{G}X}$ & $G_{\lambda\mu}^a \widetilde G^{\lambda,a}_{\;\;\nu}X^{\mu}X^{\nu}$ & $-$ & \tgray{$\mQ_{DqX^2},\mQ_{DuX^2}$} & \eqref{eq:WC:V:glgl}
    \\\midrule
    $\mO_{FF\phi^2}$ & $F_{\mu\nu} F^{\mu\nu}\phi^2$ & $-$ & $\mQ_{u\phi^2}$ & \eqref{eq:WC:S:gmgm}
    \\
    $\mO_{F \widetilde F \phi^2}$ & $F_{\mu\nu} \widetilde F^{\mu\nu}\phi^2$ & $-$ & $\mQ_{u\phi^2}$ & \eqref{eq:WC:S:gmgm}
    \\
    $\mO_{FF\chi}$ & $F_{\mu\nu} F^{\mu\nu}\chi^T C \chi$ & $-$ & $\mQ_{u\chi^2}$  & \eqref{eq:WC:F:gmgm}
    \\
    $\mO_{F\widetilde{F}\chi}$ & $F_{\mu\nu} \widetilde F^{\mu\nu} \chi^T C \chi$ & $-$ & $\mQ_{u\chi^2}$ & \eqref{eq:WC:F:gmgm}
    \\
    $\mO_{FFXX}$ & $F_{\mu\nu} F^{\mu\nu}X_{\rho}X^{\rho}$ & $-$ & $\mQ_{uX^2},\tgray{\mQ_{DqX^2},\mQ_{DuX^2}}$ & \eqref{eq:WC:V:gmgm}
    \\
    $\mO_{F\widetilde{F}XX}$ & $F_{\mu\nu} \widetilde F^{\mu\nu}X_{\rho}X^{\rho}$ & $-$ & $\mQ_{uX^2},\tgray{\mQ_{DqX^2},\mQ_{DuX^2}}$ & \eqref{eq:WC:V:gmgm}
    \\
    $\mO_{FXFX}$ & $F_{\lambda\mu} F^{\lambda}_{\;\;\nu}X^{\mu}X^{\nu}$ & $-$ & \tgray{$\mQ_{DqX^2},\mQ_{DuX^2}$} & \eqref{eq:WC:V:gmgm}
    \\
    $\mO_{FX\widetilde{F}X}$ & $F_{\lambda\mu} \widetilde F^{\lambda}_{\;\;\nu}X^{\mu}X^{\nu}$ & $-$ & \tgray{$\mQ_{DqX^2},\mQ_{DuX^2}$} & \eqref{eq:WC:V:gmgm}
    \\\bottomrule
    \end{tabular}
    \end{adjustbox}
    \caption{DLEFT operators involving scalar ($\phi$), fermionic ($\chi$), vector ($X^\mu$) DM, and ALP ($a$). The DSMEFT operators that are matched onto them at tree and one-loop levels are also shown. Operators marked in blue generate only flavour-conserving coefficients, whereas the green ones can also generate flavour-changing coefficients. Contributions from operators marked in gray vanish in the $\overline{\text{MS}}$ scheme.}
    \label{tab:DLEFT}
\end{table}

\subsection{RGE}
\label{sec:RGE}

Having specified the DLEFT operator basis, we now discuss the RG evolution of its Wilson coefficients under strong interaction. Their initial values at the EW scale $\mu_{\text{EW}}$ are obtained by matching the DSMEFT onto the DLEFT, as will be detailed in section~\ref{sec:matching}. To predict rare $B$- and $K$-meson decays, DM direct detection, and the relic abundance, these coefficients must then be evolved to the corresponding characteristic scales according to the DLEFT RG equations,
\begin{align}
    \frac{d L_i(\mu)}{d \ln \mu}=\frac{\alpha_s}{4\pi}\gamma_{ji} L_j(\mu),
\end{align}
where $L_i(\mu)$ are the Wilson coefficients of the DLEFT operators $\mO_i$ at the scale $\mu$, and $\gamma$ is the anomalous dimension matrix (ADM). Since the DM fields are colour singlets, the QCD running acts only on the quark currents. The relevant ADMs can be obtained from those of the semileptonic operators in the LEFT~\cite{Buchalla:1995vs,Jenkins:2017dyc}.

For the DLEFT operators with non-vanishing matching coefficients, the diagonal ADM entries read
\begin{numcases}{\gamma_{ii}=}
    -6C_F & $\text{for}~\mO_i=\mO_{d\phi},\mO_{d\phi^2},\mO_{dX^2},\mO_{d\chi}^{S,LR},\mO_{d\chi}^{S,RR},$ \nonumber
    \\[0.2em]
    +2C_F & $\text{for}~\mO_i=\mO_{dX}^T,\mO_{dX^2}^T,\mO_{d\chi}^{T,RR},$ \label{eq:ADM}
    \\[0.2em]
    0 & $\text{for}~\mO_i=\mO_{\phi d}^{L,R},\mO_{d\chi}^{V,LR},\mO_{d\chi}^{V,RR},\mO_{dX}^{L,R},\mO_{dXX}^{L,R},\mO_{dX\widetilde X}^{L,R},\mO_{DdX^2}^{L,R},\mO_{da}^{L,R},$ \nonumber
    \\[0.2em]
    -2\beta_0 & $\text{for}~\mO_i=\mO_{GG\phi^2},\mO_{GG\chi},\mO_{GGXX},$ \nonumber
\end{numcases}
where $\beta_0=11 N_c/3-2 N_f/3$, $C_F=(N_c^2-1)/(2N_c)$, $N_c=3$, and $N_f$ is the number of active quark flavours. For specific flavour indices, some DLEFT operators mix under QCD RG evolution. The two subspaces $\bigl([\mO_{d(u)\phi^2}]_{pp}, \mO_{GG\phi^2}\bigr)^T$ and $\bigl([\mO_{d(u)X^2}]_{pp}, \mO_{GGXX}\bigr)^T$ share the ADM submatrix
\begin{align}\label{eq:ADM:mixing:1}
    \gamma_{2 \times 2}=
    \begin{bmatrix}
        -6 C_F & 0
        \\
        \; 24 \overline{m}_p C_F\; & \; -2\beta_0 \;
    \end{bmatrix},
\end{align}
and the ADM submatrix for the subspace $\bigl([\mO_{d(u)\chi}^{S,LR}]_{pp}, [\mO_{d(u)\chi}^{S,RR}]_{pp},\mO_{GG\chi}\bigr)^T$ is given by
\begin{align}\label{eq:ADM:mixing:2}
    \gamma_{3\times 3}=
    \begin{bmatrix}
        -6 C_F & 0 & 0
        \\
        0 & -6C_F & 0
        \\
        \; 24 \overline{m}_p C_F\; & \; 24 \overline{m}_p C_F\; & \; -2\beta_0 \;
    \end{bmatrix},
\end{align}
where $m_p$ is the running mass. Both submatrices are derived from refs.~\cite{Hisano:2015rsa,Vecchi:2013iza,Hisano:2017jmz}.

We solve the RG equations numerically to obtain the evolution factors for rare $B$ ($\mu_B=4.8\GeV$) and $K$ ($\mu_K=2.0\GeV$) decays:
\begin{align}
    L_i(\mu_B)&=L_i(m_Z)\cdot
\begin{cases}
    1.37&\text{for}~\mO_i=\mO_{d\phi},\mO_{d\phi^2},\mO_{dX^2},\mO_{d\chi}^{S,LR},\mO_{d\chi}^{S,RR},
    \\
    0.90&\text{for}~\mO_i=\mO_{dX}^T,\mO_{dX^2}^T,\mO_{d\chi}^{T,RR},
\end{cases}\\
    L_i(\mu_K)&=L_i(m_Z)\cdot
\begin{cases}
    1.61&\text{for}~\mO_i=\mO_{d\phi},\mO_{d\phi^2},\mO_{dX^2},\mO_{d\chi}^{S,LR},\mO_{d\chi}^{S,RR},
    \\
    0.85&\text{for}~\mO_i=\mO_{dX}^T,\mO_{dX^2}^T,\mO_{d\chi}^{T,RR}.
    \end{cases}
\end{align}
For operators that mix under QCD RG evolution, scalar and vector DM have identical evolution matrices. For scalar DM, the numerical results at $\mu_K$ are given by
\begin{align}\label{eq:QCD-mixing-scalar-muK}
    \begin{pmatrix}
        [L_{u\phi^2}]_{11}\\
        [L_{u\phi^2}]_{22}\\
        [L_{d\phi^2}]_{11}\\
        [L_{d\phi^2}]_{22}\\
        L_{GG\phi^2}
    \end{pmatrix}(\mu_K)
    =\begin{pmatrix}
        1.61 & 0 & 0 & 0 & 0.99 \\
        0 & 1.61 & 0 & 0 & 0.09\\
        0 & 0 & 1.61 & 0 & 0.99\\
        0 & 0 & 0 & 1.61 & 0.84\\
        0 & 0 &  0 &  0 & 2.70
    \end{pmatrix}\begin{pmatrix}
        [L_{u\phi^2}]_{11}\\
        [L_{u\phi^2}]_{22}\\
        [L_{d\phi^2}]_{11}\\
        [L_{d\phi^2}]_{22}\\
        L_{GG\phi^2}
    \end{pmatrix}(m_Z).
\end{align}
The corresponding evolution for vector DM is obtained by replacing $\phi$ with $X$. For fermionic DM, the numerical results at $\mu_K$ read
\begin{align}\label{eq:QCD-mixing-fermion-muK}
     \begin{pmatrix}
        [L_{u\chi}^{S,\,LR}]_{11}\\
        [L_{u\chi}^{S,\,LR}]_{22}\\
        [L_{d\chi}^{S,\,LR}]_{11}\\
        [L_{d\chi}^{S,\,LR}]_{22}\\
        L_{GG\chi}
    \end{pmatrix}(\mu_K)
    =\begin{pmatrix}
        1.61 & 0 & 0 & 0 & 0.99\\
        0 & 1.61 & 0 & 0 & 0.09 \\
        0 & 0 & 1.61 & 0 & 0.99\\
        0 & 0 & 0 & 1.61 & 0.84\\
        0 & 0 & 0 &  0 &  2.70
    \end{pmatrix}\begin{pmatrix}
        [L_{u\chi}^{S,\,LR}]_{11}\\
        [L_{u\chi}^{S,\,LR}]_{22}\\
        [L_{d\chi}^{S,\,LR}]_{11}\\
        [L_{d\chi}^{S,\,LR}]_{22}\\
        L_{GG\chi}
    \end{pmatrix}(m_Z).
\end{align}
The $RR$ coefficients obey the same evolution as the $LR$ coefficients, and there is no mixing between these two sectors. These results can also be applied to the analysis of DM relic density. For DM direct detection, which probes DM--nucleon interactions, we evolve the Wilson coefficients to $\mu_H=1.0\GeV$. The corresponding numerical results are given by
\begin{align}
    L_i(\mu_H)&=L_i(m_Z)\cdot
\begin{cases}
    1.97&\text{for}~\mO_i=\mO_{d\phi},\mO_{d\phi^2},\mO_{dX^2},\mO_{d\chi}^{S,LR},\mO_{d\chi}^{S,RR},
    \\
    0.80&\text{for}~\mO_i=\mO_{dX}^T,\mO_{dX^2}^T,\mO_{d\chi}^{T,RR}.
    \end{cases}
\end{align}
For operators that mix under QCD RG evolution, we present only the scalar-DM case and the $LR$ sector for fermionic DM, as in eqs.~\eqref{eq:QCD-mixing-scalar-muK} and~\eqref{eq:QCD-mixing-fermion-muK}. Numerically, we have at the scale $\mu_H$ 
\begin{align}
    \begin{pmatrix}
        [L_{u\phi^2}]_{11}\\
        [L_{d\phi^2}]_{11}\\
        [L_{d\phi^2}]_{22}\\
        L_{GG\phi^2}
    \end{pmatrix}(\mu_H)&=\begin{pmatrix}
        1.97 & 0 & 0 & 0.99\\
        0 & 1.97 & 0 & 0.99\\
        0 & 0 & 1.97 & 0.78\\
        0 & 0 & 0 & 4.61
    \end{pmatrix}\begin{pmatrix}
        [L_{u\phi^2}]_{11}\\
        [L_{d\phi^2}]_{11}\\
        [L_{d\phi^2}]_{22}\\
        L_{GG\phi^2}
    \end{pmatrix}(m_Z), \nonumber\\[1em]
     \begin{pmatrix}
        [L_{u\chi}^{S,\,LR}]_{11}\\
        [L_{d\chi}^{S,\,LR}]_{11}\\
        [L_{d\chi}^{S,\,LR}]_{22}\\
        L_{GG\chi}
    \end{pmatrix}(\mu_H)
    &=\begin{pmatrix}
        1.97 & 0 & 0 & 0.99\\
        0 & 1.97 & 0 & 0.99\\
        0 & 0 & 1.97 & 0.78\\
        0 & 0 & 0 & 4.61
    \end{pmatrix}\begin{pmatrix}
        [L_{u\chi}^{S,\,LR}]_{11}\\
        [L_{d\chi}^{S,\,LR}]_{11}\\
        [L_{d\chi}^{S,\,LR}]_{22}\\
        L_{GG\chi}
    \end{pmatrix}(m_Z).
\end{align}

Since the operators $\mO_{G\widetilde G \phi^2}$, $\mO_{G \widetilde G \chi}$, $\mO_{G \widetilde G X X}$, $\mO_{GXGX}$, and $\mO_{GX \widetilde G X}$ do not enter our phenomenological analyses, we do not give here the numerical evolution matrices corresponding to them.

\section{Matching between DSMEFT and DLEFT}
\label{sec:matching}

In this section, the DLEFT is considered as the low-energy limit of the DSMEFT and we perform the one-loop matching between the two EFTs at the EW scale. Since our analysis focuses on light DM, the DM fields are kept as active low-energy degrees of freedom, while heavy SM particles with masses at or above the EW scale are all integrated out. The matching is carried out using the standard off-shell matching approach~\cite{Georgi:1993mps,Manohar:2018aog}.

\begin{figure}[t]
    \centering
    \subfloat[]{
      \includegraphics[width=0.25\linewidth]{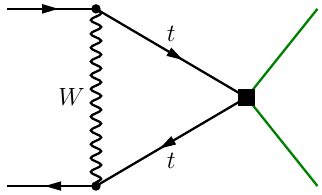}
      \label{fig:feynman diagram:prDMDM}
    }
    \qquad
    \subfloat[]{
      \includegraphics[width=0.25\linewidth]{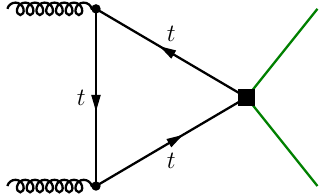}
      \label{fig:feynman diagram:glglDMDM}
    }
    \qquad
    \subfloat[]{
      \includegraphics[width=0.25\linewidth]{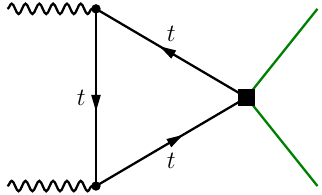}
      \label{fig:feynman diagram:gmgmDMDM}
    }
    \\
    \subfloat[]{
      \includegraphics[width=0.25\linewidth]{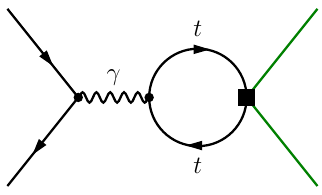}
      \label{fig:feynman diagram:ppDMDM:1} 
    }
    \qquad\hspace{0.075em}
    \subfloat[]{
      \includegraphics[width=0.25\linewidth]{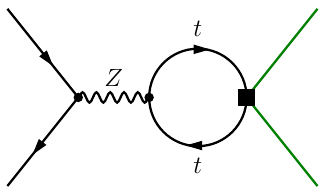}
      \label{fig:feynman diagram:ppDMDM:2} 
    }
    \qquad
    \subfloat[]{
      \includegraphics[width=0.25\linewidth]{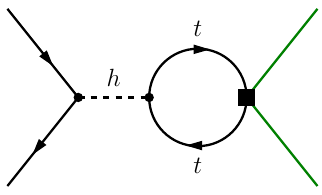}
      \label{fig:feynman diagram:ppDMDM:3} 
    }
  \caption{Feynman diagrams involved in the top-flavoured DSMEFT during the one-loop matching in unitary gauge. The green lines represent the DM particles, and the DSMEFT operator is represented by the black square.}
  \label{fig:feynman diagram:one-loop matching}
\end{figure}

The relevant one-loop Feynman diagrams are shown in figure~\ref{fig:feynman diagram:one-loop matching} and can be divided into two categories. The first category contains the diagrams shown in figures~\ref{fig:feynman diagram:prDMDM}, \ref{fig:feynman diagram:ppDMDM:1}, \ref{fig:feynman diagram:ppDMDM:2}, and \ref{fig:feynman diagram:ppDMDM:3}. For these diagrams, the calculation can be performed within the naive dimensional regularization (NDR) scheme~\cite{Buras:1989xd,Buras:2011we}, irrespective of the operator inserted. The second category consists of the diagrams shown in figures~\ref{fig:feynman diagram:glglDMDM} and \ref{fig:feynman diagram:gmgmDMDM}, in which two gauge bosons are attached to a closed fermion loop and the EFT vertex may contain an insertion of an axial-vector current. Consequently, traces such as $\operatorname{tr}(\gamma_5\gamma^\mu\gamma^\nu\gamma^\rho\gamma^\sigma)$ arise, and a naive application of the NDR scheme would violate the Ward identities. For these diagrams, we adopt the ``reading point'' prescription~\cite{Korner:1991sx,Kreimer:1993bh}: (i) the starting point for reading the fermion loop is fixed when constructing the amplitude; and (ii) only the algebraic properties of the $\gamma$ matrices are used, without exploiting the cyclicity of the trace. We have explicitly verified that the resulting amplitudes satisfy the Ward identities and are dependent of the choice of reading point.

The calculation is performed with the aid of several public packages, including \texttt{FeynRules}~\cite{Alloul:2013bka}, \texttt{FeynArts}~\cite{Hahn:2000kx}, \texttt{FeynCalc}~\cite{Mertig:1990an,Shtabovenko:2016sxi,Shtabovenko:2020gxv}, and \texttt{Package-X}~\cite{Patel:2016fam}, supplemented by in-house routines. For DSMEFT operators involving left-handed vector currents built from quark doublets (e.g., $\mQ_{q\chi}$ in table~\ref{tab:DSMEFT}), we include only the tree-level matching contribution, since these operators already contribute to the DLEFT at tree level. A summary of the tree- and one-loop matching contributions is provided in table~\ref{tab:DLEFT}, while the explicit matching coefficients are relegated in appendix~\ref{app:matching coefficients}.

As a cross-check, we evaluate the diagrams in both the Feynman and unitary gauges, obtaining completely consistent results. In addition, for the matching coefficients of the operators $\mO_{GG\phi^2}$, $\mO_{G\widetilde G\phi^2}$, $\mO_{GG\chi}$, $\mO_{G\widetilde G\chi}$, $\mO_{GGXX}$, and $\mO_{G\widetilde GXX}$, we find agreement with the results obtained by directly applying the trace anomaly of the energy-momentum tensor~\cite{Shifman:1978zn},
\begin{align}\label{eq:anomaly_matching}
    m_t\bar t t \to -\frac{\alpha_s}{12\pi}G^a_{\mu\nu}G^{\mu\nu,a},
    \qquad
    m_t\bar t i\gamma^5 t \to \frac{\alpha_s}{8\pi}G^a_{\mu\nu}\widetilde G^{\mu\nu,a}.
\end{align}

\section{Flavour observables}
\label{sec:flavour}

The matching of the top-flavoured DSMEFT operators onto the DLEFT, derived in section~\ref{sec:matching}, generates operators that mediate the decays $M_1\to M_2+\DM\,(+\DM)$ of $B$ and $K$ mesons. Since the DM particles escape detection, these modes are experimentally indistinguishable from the decay $M_1\to M_2\nu\bar\nu$, and these two contributions add incoherently,
\begin{align}\label{eq:BR:inv}
    \mB(M_1\to M_2+\inv)=\mB_{\SM}(M_1\to M_2\nu\bar\nu)+\mB_{\NP}\big(M_1\to M_2+\DM +\DM \big).
\end{align}
The same operators also mediate the invisible decays $B_{s,d}\to\DM + \DM$, the SM rates of which are negligibly small. We summarize in table~\ref{tab:observables} the $b\to s$, $b\to d$ and $s\to d$ modes considered in our analysis, together with their SM predictions, current experimental bounds and projected sensitivities. The NP amplitudes factor into the DLEFT Wilson coefficients and the hadronic matrix elements. The Wilson coefficients are evolved to $\mu_B$ or $\mu_K$ as described in section~\ref{sec:RGE}, while we use the form factors and decay constants employed in ref.~\cite{Hou:2024vyw}. The decay rates for $d_i\to d_j+\phi\,(+\phi)$~\cite{Kamenik:2011vy,Bird:2004ts,Li:2019cbk,He:2020jly,Geng:2020seh,Li:2021sqe,Kling:2022uzy}, $d_i\to d_j+\chi\bar\chi$~\cite{Kamenik:2011vy,Felkl:2021uxi,Su:2019tjn,Li:2020dpc}, $d_i\to d_j+X\,(+X)$~\cite{Kamenik:2011vy,He:2022ljo,Li:2021sqe} and $d_i\to d_j+a$~\cite{MartinCamalich:2020dfe,Bauer:2021mvw,Guerrera:2022ykl} are evaluated with the package \texttt{HadronToNP}~\cite{HadronToNP:2024}, following the same procedure as described in ref.~\cite{Hou:2024vyw}.

\begin{table}[t]
    \renewcommand*{\arraystretch}{1.3}
	\tabcolsep 0.16in
    \centering
    \begin{tabular}{c| c |c|c|c}
      \toprule
      Observable    &  SM   & Exp. & Unit & Sensitivity  \\
      \midrule
      $\mB(B^+ \to K^+ \nu \bar{\nu}) $   & $4.19 \pm 0.57$  &  $23 \pm 7$ \cite{Belle-II:2023esi} &$10^{-6} $  & $19\%$ \cite{Belle-II:2022cgf}  \\
      $\mB(B^0 \to K^0 \nu \bar{\nu})$  & $3.87 \pm 0.51$ &  $<26 $ \cite{Belle:2017oht}& $ 10^{-6}$ & $87\% $ \cite{Belle-II:2022cgf}   \\
      $\mB(B^+ \to K^{*+} \nu \bar{\nu})$ & $9.75 \pm 0.91$&  $<61 $ \cite{Belle:2017oht}&$ 10^{-6} $ & $75\%$ \cite{Belle-II:2022cgf}\\
      $\mB(B^0 \to K^{*0} \nu \bar{\nu}) $ & $9.00 \pm 0.84$ & $<18 $ \cite{Belle:2017oht}&$ 10^{-6}$ & $40\%$ \cite{Belle-II:2022cgf}\\
      $\mB(B_s \to \phi_s \nu \bar\nu)$~${}^a$  & $9.68\pm 0.73$ & $<5400$ \cite{DELPHI:1996ohp}& $10^{-6}$ & $2\%$ \cite{Li:2022tov}\\
      \midrule
      $\mB(B^+ \to \pi^+ \nu \bar{\nu}) $ & $1.29 \pm 0.18$&  $<140$ \cite{Belle:2017oht}&$ 10^{-7} $ &  \\
      $\mB(B^0 \to \pi^0 \nu \bar{\nu}) $ & $6.00 \pm 0.77$ &  $<900$  \cite{Belle:2017oht}&$ 10^{-8} $ &  \\
      $\mB(B^+ \to \rho^+ \nu \bar{\nu}) $ & $3.73 \pm 0.73$ &  $<300$  \cite{Belle:2017oht}&$ 10^{-7} $ & \\
      $\mB(B^0 \to \rho^0 \nu \bar{\nu}) $ & $1.73 \pm 0.36$ &   $<400$ \cite{Belle:2017oht}&$ 10^{-7} $ & \\
      \midrule
      $\mB(K^+ \to  \pi^+ \nu \bar{\nu}) $  & $8.00 \pm 0.60$ &  $9.6 ^{+1.9}_{-1.8}$ \cite{Chang:2026vvx} & $10^{-11} $ &   \\
      $\mB(K_L \to  \pi^0 \nu \bar{\nu}) $  & $3.13 \pm 0.33$  &  $<2.2 $ \cite{KOTO:2024zbl} &$ 10^{-9} $ &  \\
      \midrule
      $\mB(B_s \to \nu \bar{\nu})$  & $\approx 0$ & $<5.9$ \cite{Alonso-Alvarez:2023mgc}& $10^{-4}$ & $1.1 \times 10^{-5}$ \cite{Belle-II:2018jsg}
      \\
      $\mB(B^0 \to \nu \bar{\nu})$  & $\approx 0$ & $<1.4$ \cite{Alonso-Alvarez:2023mgc}& $10^{-4}$ & $5.0 \times 10^{-6}$ \cite{Belle-II:2018jsg}
      \\
      \bottomrule
    \end{tabular}
    \caption{SM predictions and experimental results for the flavour observables considered. All upper limits are given at the $90\%$ confidence level (CL). The last column shows the projected sensitivities (i.e., the expected relative uncertainty) for the $\nu\bar\nu$ modes, as well as the expected upper limit for the invisible modes at Belle~II with $5\ab^{-1}$ and, for $B_s\to\phi_s\nu\bar\nu$, at the Tera-$Z$ phase of the CEPC. The projection for the FCC-ee is comparable to that for the CEPC and is not listed separately. ${}^a$ To distinguish the meson $\phi$ from the scalar DM $\phi$, we have renamed the meson $\phi$ by $\phi_s$.}
    \label{tab:observables}
\end{table}

In the top-flavoured scenario, all down-type FCNC transitions induced by a given DSMEFT operator are controlled by a single coupling $[\mC_i]_{33}$. These transitions are mediated by the DLEFT operators whose Wilson coefficients are fixed by the matching conditions derived in section~\ref{sec:matching}. The explicit expressions collected in appendix~\ref{app:matching coefficients} reveal three generic features of the top-flavoured DM contributions to these processes. All of them can be read off from the matching results for the three representative DSMEFT operators,
\begin{align}\label{eq:flavour:structure}
    \big[L_{d\phi^2}\big]_{pr} &= V_{tp}^*V_{tr}\,\frac{m_r\,m_t}{16\sqrt2\pi^2\Lambda^2}\Big ([\mC_{u\phi^2}]_{33}\,S_1(x_t)+[\mC_{u\phi^2}]_{33}^*\,S_2(x_t)\Big ),
    \nn\\[0.2em]
    \big[L_{\phi d}^{L}\big]_{pr} &= V_{tp}^*V_{tr}\left (\frac{v^2}{\Lambda^2}\,[\mC_{\phi q}]_{33}+\frac{m_t^2}{16\pi^2\Lambda^2}\,S_3(x_t)\,[\mC_{\phi u}]_{33}\right ),
\end{align}
with $x_t=m_t^2/m_W^2$. First, both of the two DLEFT coefficients carry the CKM factor $V_{tp}^*V_{tr}$, which coincides with that of the top-quark loops in the SM, as in the MFV scenario beyond the SM~\cite{DAmbrosio:2002vsn}. The $b\to s$, $b\to d$ and $s\to d$ amplitudes are therefore correlated with each other, scaling naively as $|V_{tb}V_{ts}^*|:|V_{tb}V_{td}^*|:|V_{ts}V_{td}^*|\simeq 1:0.21:0.01$. The rare $K$-meson decays nevertheless remain competitive, because the experimental uncertainty of $\mB(K^+\to\pi^+\nu\bar\nu)$ is several orders of magnitude smaller than those of rare $B$-meson decays. Second, the DLEFT coefficients of the chirality-flipped operators $\mQ_{u\phi}$, $\mQ_{u\phi^2}$, $\mQ_{u\chi^2}$ and $\mQ_{uX^2}$ are proportional to $m_r$, the mass of the external right-handed down-type quark $d_r$, as is evident from the first line of eq.~\eqref{eq:flavour:structure} for $\mQ_{u\phi^2}$. This mass factor is $\overline{m}_b$ in $b\to s,d$ and $\overline{m}_s$ in $s\to d$ transitions, making the kaon decay amplitudes suppressed by a further factor $\overline{m}_s/\overline{m}_b\simeq0.02$. For $B_q\to\DM + \DM$, on the contrary, the factor $\overline{m}_b$ is compensated by the pseudoscalar matrix element $\langle0|\bar q\gamma_5 b|\bar{B}_q\rangle \propto f_{B_q}m_{B_q}^2/(\overline{m}_b+\overline{m}_q)$, so that, unlike the SM axial-vector-current contributions, these decays are not suppressed by helicity. Third, as shown by the second line of eq.~\eqref{eq:flavour:structure} for $\mQ_{\phi q}$ and $\mQ_{\phi u}$, each operator with a quark-doublet current $\bar q\gamma^\mu q$ and its partner with an up-type singlet current $\bar u\gamma^\mu u$ match onto the same DLEFT operator, with the former contributing at tree and the latter at one-loop level, respectively. Since the matching coefficients do not depend on the DM mass, the two operators of each pair yield rates with identical dependence on $m_{\DM}$, and the bounds on the singlet operators are weaker than those on their doublet partners by the universal ratio of the tree-level to the one-loop prefactor, $16\pi^2v^2/(m_t^2|S_3(x_t)|)\simeq 210$. A similar observation with a mild variation also applies for the DSMEFT operator pair $\mQ_{DqX^2}$ and $\mQ_{DuX^2}$: the loop-induced contribution from $\mQ_{DuX^2}$ involves a different loop function, leading to a rescaling factor of about $1200$, and both operators also match onto the same DLEFT operator $\mO_{dX^2}$. Thus, we show results only for the doublet operators; the corresponding bounds on the singlet operators can be obtained by appropriate rescaling.

Having discussed the generic structure of the NP contributions, we now turn to our numerical analysis. The decays into two DM particles, the decays into a single DM particle, and the prospects for future measurements are discussed in turn. Throughout our analysis, only one DSMEFT operator is switched on at a time, with a real Wilson coefficient and $\Lambda=1\TeV$.\footnote{In principle, the NP scale $\Lambda$ should be fixed at the EW scale to be consistent with our previous assumption made in footnote~\footref{fn:DSMEFTatmuEW}. A simple rescaling is needed to obtain the same Wilson coefficient $\mC^\prime$ at another scale $\Lambda^\prime$. Our choice of $\Lambda=1\TeV$ is just for ease of this rescaling.} To determine the viable range of each Wilson coefficient, the Belle~II measurement~\cite{Belle-II:2023esi} of $\mB(B^+\to K^+\nu\bar\nu)$ is imposed at the $2\sigma$ level, while the other observables in table~\ref{tab:observables} are imposed at $90\%$~CL.

\begin{figure}[htbp]
\centering
\includegraphics[width=0.96\linewidth]{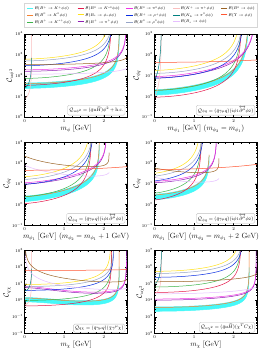}
\caption{Flavour constraints on the Wilson coefficients of the DSMEFT operators with two scalar (upper four) or two fermionic (lower two) DM fields as functions of the DM mass, at the NP scale $\Lambda=1\TeV$. The band is allowed by the Belle~II measurement of $\mB(B^+\to K^+\nu\bar\nu)$ at the $2\sigma$ level, and the solid lines correspond to the $90\%$~CL upper bounds from the other observables in table~\ref{tab:observables}.}
\label{fig:flavour:two}
\end{figure}

\begin{figure}[t]
    \centering
    \includegraphics[width=0.96\linewidth]{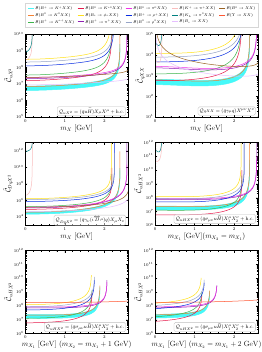}
    \caption{Same as in figure~\ref{fig:flavour:two}, but for the DSMEFT operators with two vector DM fields.}
    \label{fig:flavour:two:vector}
\end{figure}

For the decays into two DM particles, we focus on the eight DSMEFT operators $\mQ_{\phi q}$, $\mQ_{q\chi}$, $\mQ_{u\phi^2}$, $\mQ_{u\chi^2}$, $\mQ_{uX^2}$, $\mQ_{qXX}$, $\mQ_{DqX^2}$, and $\mQ_{uHX^2}$ that give independent constraining patterns in these modes. As established above, the operators not displayed separately introduce no distinct dependence on the DM mass, since their effects can be related to those of the representative operators by an overall rescaling. For example, the contributions induced by $\mQ_{qXX}$, $\mQ_{uXX}$, $\mQ_{q\widetilde XX}$ and $\mQ_{u\widetilde XX}$ have identical dependence on $m_X$ and differ only in their overall normalization. The corresponding flavour constraints on these representative operators are shown in figures~\ref{fig:flavour:two} and \ref{fig:flavour:two:vector}. The range of the Wilson coefficients allowed by the Belle~II measurement of $\mB(B^+\to K^+\nu\bar\nu)$ at the $2\sigma$ level, which is referred to as the Belle~II band in the following, ends at the kinematic endpoint $m_{\DM}=(m_B-m_K)/2 \simeq 2.4\GeV$, towards which the required Wilson coefficient grows steeply. The pattern of constraints is governed mainly by the Lorentz structure of the quark current in the induced DLEFT operators and by the spin of the DM. We therefore divide the operators into the following four groups and discuss them one by one.
\begin{itemize}

\item For the vector-current operators $\mQ_{\phi q}$ and $\mQ_{q\chi}$, the Belle~II excess in $\mB(B^+\to K^+\nu\bar\nu)$ requires $\mathcal O(1)$ coefficients. For DM masses below about $1.6\GeV$, most of the Belle~II band is excluded by $\mB(B^0\to K^{*0}+\inv)$. This can be understood from the fact that the induced left-handed quark current has the same structure as in the SM and couples more strongly to the $K^*$ mode than to the $K$ mode. Indeed, we have $\mB(B^0\to K^{*0}\nu\bar\nu)/\mB(B^+\to K^+\nu\bar\nu)\simeq 2.2$ in the SM, while the current experimental upper limit on $\mB(B^0\to K^{*0}\nu\bar\nu)$ is only about twice its SM prediction. For larger DM masses, the $B\to K^*$ phase space shuts off sooner than the $B\to K$ channel, which relaxes this constraint and allows viable parameter ranges to emerge. For the $\mQ_{q\chi}$ operator, the upper end of the viable window is set by $\mB(B^+\to\pi^+ \chi \bar\chi)$. For the $\mQ_{\phi q}$ operator with non-degenerate scalars satisfying $m_{\phi_2}-m_{\phi_1}=1\,(2)\GeV$, the window for the lighter state moves to $1.1$--$1.8\GeV$ ($0.1$--$1.3\GeV$). Its upper end is set by $\mB(B_s\to\phi_1\phi_2)$, whose amplitude is proportional to $m_{\phi_2}^2-m_{\phi_1}^2$ and vanishes in the degenerate case. The decay $\Upsilon(1S)\to\inv$ directly constrains the flavour-diagonal $\bar{b}b$ couplings~\cite{BaBar:2009gco,Kumar:2013iva,Fernandez:2014eja}, which are related to $[\mC_i]_{33}$ by the unsuppressed CKM factor $|V_{tb}|^2\simeq1$. For the $\mQ_{\phi q}$ operator in the degenerate-scalar case, the upper end of the viable window is set by this decay.

\item For the scalar-current operators $\mQ_{u\phi^2}$, $\mQ_{u\chi^2}$ and $\mQ_{uX^2}$, the $B\to K^*$ transition proceeds only through the form factor $A_0$ and is therefore less constraining. For $\mQ_{u\phi^2}$, despite the $\overline{m}_s/\overline{m}_b$ suppression, $\mB(K^+\to\pi^++\inv)$ provides the strongest bound for light DM and excludes the Belle~II band for $m_\phi \lesssim 0.2\GeV$. At larger masses, $\mB(B^0\to K^{*0}+\inv)$ removes the upper part of the band up to $m_\phi\simeq1.4\GeV$, and the unsuppressed decay $B_s\to\phi\phi$ excludes the band for $m_\phi>2.2\GeV$. For $\mQ_{u\chi^2}$ and $\mQ_{uX^2}$, the Belle~II band is essentially unconstrained below about $2\GeV$ by other observables, apart from a small portion excluded by $\mB(B^0\to K^0+\inv)$. In the high-mass region, $\mQ_{uX^2}$ behaves like $\mQ_{u\phi^2}$, with a large part of the band excluded by $\mB(B_s\to\inv)$, whereas for $\mQ_{u\chi^2}$ most of the band is excluded by $\mB(B^+\to\pi^++\inv)$.

\item The remaining vector-DM operators, $\mQ_{qXX}$ and $\mQ_{DqX^2}$, contain a left-handed quark current. For light DM, $\mB(B^0\to K^{(*)0}XX)$ exclude only a small portion of the Belle~II band close to its upper edge, whereas $\mB(B_s\to XX)$ becomes the dominant constraint at larger masses. The reach of the latter depends on the Lorentz structure. For $\mQ_{DqX^2}$, the $B_s\to XX$ rate is enhanced by the longitudinal polarization of $X$ in the same way as for the $B\to KXX$ rate, so that the bound on $\tilde\mC_{DqX^2}$ depends only mildly on $m_X$: it enters the band at $m_X\simeq1.0\GeV$ and excludes it for $m_X>1.8\GeV$, the lowest mass among all operators with two identical DM fields. For $\mQ_{qXX}$, this enhancement is absent in the two-body decay, and the bound on $\tilde\mC_{qXX}$ scales as $1/m_X$, making it irrelevant for light $X$ but excluding the band for $m_X\gtrsim2\GeV$.

\item The operator $\mQ_{uHX^2}$ behaves similarly to $\mQ_{\phi q}$ in the low-mass region, where it is strongly constrained by $\mB(B^0\to K^{*0}+\inv)$. Specifically, for $m_{X_2}-m_{X_1}=0$, $1$ and $2\GeV$, the ranges with $m_{X_1}\lesssim 2$, $1.5$ and $1\GeV$ are excluded, respectively. Unlike $\mQ_{\phi q}$, however, $\mQ_{uHX^2}$ matches onto the tensor DLEFT operator $\mO_{dX^2}^T$. Since the hadronic matrix element $\langle0|\bar q\sigma_{\mu\nu}b|\bar{B}_q\rangle$ vanishes, this operator does not contribute to $B_q\to\inv$. Its high-mass range is instead constrained by $\mB(B^+\to\pi^++\inv)$.
\end{itemize}
We note that only the tree-level operators can accommodate the Belle~II excess with $\mathcal O(1)$ coefficients at the TeV scale. The loop-induced operators generally require coefficients of $\mathcal O(10$--$100)$, e.g., $\mC_{u\phi^2}\simeq25$--$45$.\footnote{The only exception is $\mQ_{u\chi^2}$, which requires $\mC_{u\chi^2}=\mathcal O(10^3)$ because it is a dimension-7 operator.} This would imply strongly coupled NP close to the EW scale, at the border of validity of the EFT.

\begin{figure}[t]
    \centering
    \includegraphics[width=0.96\linewidth]{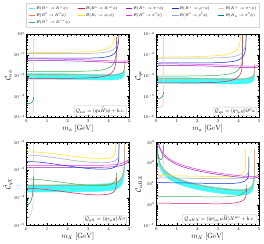}
    \caption{Same as in figure~\ref{fig:flavour:two}, but for the DSMEFT operators with a single DM field.}
    \label{fig:flavour:one}
\end{figure}

For the two-body decays $B\to K^{(*)}+\DM$, we focus on the four DSMEFT operators that give independent constraining patterns in these channels, namely $\mQ_{u\phi}$, $\mQ_{qa}$, $\mQ_{qX}$, and $\mQ_{uHX}$. As discussed above, other single-DM operators lead to the same DM-mass dependence up to a simple rescaling of the Wilson coefficient; for example, the contributions induced by $\mQ_{uX}$, $\mQ_{HqX}^{(1,3)}$ and $\mQ_{HuX}$ differ from those induced by $\mQ_{qX}$ only by an overall factor. The results are shown in figure~\ref{fig:flavour:one}. Here we apply the Belle~II result directly to $B^+\to K^++\DM$, although the experimental efficiency depends on the $q^2$ distribution, which for a two-body decay is a monochromatic peak at $q^2=m_{\DM}^2$; a dedicated reinterpretation, as in ref.~\cite{Altmannshofer:2023hkn}, is beyond the scope of this work. For $m_{\DM} \lesssim m_K-m_\pi\simeq0.35\GeV$, the bound from $\mB(K^+\to\pi^++\inv)$ lies at least an order of magnitude below the Belle~II band, making this range excluded for all operators. At larger masses, the decisive constraint is again from $\mB(B^0\to K^{*0}+\inv)$. For $\mQ_{u\phi}$ and $\mQ_{qa}$, it runs just above the lower edge of the band, so that only a thin strip survives up to $m_{\DM}\simeq3\GeV$ and most of the band is recovered only for $m_{\DM}\gtrsim4\GeV$. For the vector-DM operator $\mQ_{qX}$, the $B\to K^*X$ rate receives a contribution from the transversely polarized $X$ that grows with $m_X$, and the band is excluded for $1.9<m_X<4.3\GeV$. The dipole operator $\mQ_{uHX}$ is the most constrained one. For a light $X$, the $B\to KX$ decay rate vanishes proportionally to $m_X^2$, whereas the $B\to K^*X$ rate remains finite; thus explaining the Belle~II excess would need a large coefficient, but this is already ruled out by the $\mB(B^0\to K^{*0}+\inv)$ constraint. In fact, this bound lies below the preferred band for all $m_X<4.4\GeV$, and only the window $4.4$--$4.8\GeV$ survives, where the decay $B\to K^*X$ is kinematically forbidden.

\begin{figure}[t]
    \centering
    \includegraphics[width=0.96\linewidth]{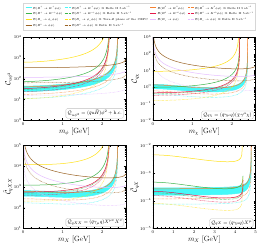}
    \caption{Projected sensitivities of the $b\to s$ observables for some representative DSMEFT operators. The dotted lines show the projected $90\%$~CL bounds at Belle~II with $5\ab^{-1}$ and, for $B_s\to\phi_s\nu\bar\nu$, at the Tera-$Z$ phase of the CEPC, assuming the SM-like central values. The band and the solid lines are the same as in figures~\ref{fig:flavour:two}--\ref{fig:flavour:one}.}
    \label{fig:flavour:proj}
  \end{figure}

The precision of the $b\to s$ measurements discussed above is expected to improve significantly in the future~\cite{Belle-II:2018jsg,Belle-II:2022cgf,Li:2022tov,Amhis:2023mpj}. The projected sensitivities at Belle~II with $5\ab^{-1}$ and during the Tera-$Z$ phase of the CEPC and FCC-ee, summarized in table~\ref{tab:observables}, are shown in figure~\ref{fig:flavour:proj} for some representative operators, with the complete results collected in figures~\ref{fig:flavour:two:b2s}--\ref{fig:flavour:one:b2s} in appendix~\ref{app:projections}. With an integrated luminosity of $5\ab^{-1}$, Belle~II is expected to measure $\mB(B^+\to K^+\nu\bar\nu)$ with a relative uncertainty of $19\%$. For an SM-like central value, this projected sensitivity would provide a particularly stringent test of the Belle~II band: the resulting bounds would lie at about half of its lower edge across all operators and DM masses, thus excluding the entire band. If instead the central value of the present measurement persists, the projected measurement of $(23\pm4)\times10^{-6}$ would deviate from the SM prediction by about $4\sigma$. The projected bound from $\mB(B^0\to K^0\nu\bar\nu)$ would also lie below the band, at about $0.74$ times its lower edge, providing an independent test. In addition to the $B\to K^{(*)}$ modes, the $B_s$ decays provide complementary probes. A measurement of $\mB(B_s\to\phi_s\nu\bar\nu)$ with a $2\,\%$ sensitivity during the Tera-$Z$ phase of the CEPC or FCC-ee would cover the Belle~II band up to $m_{\DM}\simeq2\GeV$ for most two-DM operators and up to $4\GeV$ for all single-DM operators, including the four representative cases shown in figure~\ref{fig:flavour:proj}. The projected Belle~II sensitivity of $\mB(B_s\to\inv)$ would exclude the band above $m_{\DM}\simeq2\GeV$ for most operators and over the entire DM-mass range for $\mQ_{u\phi^2}$, $\mQ_{uX^2}$ and $\mQ_{DqX^2}$. Taken together, these two $B_s$ decay modes would exclude the Belle~II band for most of the operators considered.

\section{Direct detection}
\label{sec:direct detection}

Direct-detection experiments use different targets to determine the precise interactions of a DM particle with nuclei. This is achieved by looking for nuclear recoils induced by the scattering of halo DM particles off the target nuclei. In the following subsections, we introduce the NREFT for describing the DM--nucleon interactions~\cite{Fan:2010gt,Fitzpatrick:2012ix} and detail the calculations relevant for the DM--nucleus scattering~\cite{DelNobile:2021wmp}.

\subsection{Non-relativistic EFT}

In direct-detection experiments, the interactions between DM and nuclei can be described within the NREFT framework~\cite{Fan:2010gt,Fitzpatrick:2012ix}. The most general Galilean-invariant operators are constructed from the following fundamental building blocks~\cite{Dent:2015zpa,Catena:2019hzw,Gondolo:2020wge}:
\begin{align}\label{eq:building-blocks}
    \mathds{1}_N, \qquad \mathds{1}_{\DM}, \qquad \frac{i\bq}{m_N}, \qquad \vN, \qquad \SN, \qquad \SDM, \qquad \boldsymbol{\mS},
\end{align}
where $\mathds{1}_{\DM}$ ($\mathds{1}_N$) and $\SDM$ ($\SN$) denote the identity and spin operators that act on the DM (nucleon) state space. The three-momentum transfer is given by $\bq=\bp-\bp'$, where $\bp$ ($\bp'$) denotes the incoming (outgoing) three-momentum of the DM particle. The transverse relative velocity is defined as $\bv_N^\perp\equiv\bv-\bq/(2\mu)$, where $\bv$ is the relative velocity between the incoming DM particle and the nucleon and $\mu$ is the corresponding reduced mass of the DM-nucleon system; by construction, it satisfies $\bv_N^\perp\cdot\bq=0$ for elastic scattering. Following ref.~\cite{Liang:2025kkl}, we define for vector DM the operator $\boldsymbol{\mS}$ (denoted there by $\boldsymbol{\tilde\mS}$) as a traceless symmetric tensor operator. Its components $S_X^{ij}$ are constructed from the vector-DM spin operators $S_X^i$ according to
\begin{align}
    S_X^{ij}=\frac12\big(S_X^i S_X^j+S_X^j S_X^i\big)-\frac23\delta^{ij},
\end{align}
where the subscript $X$ labels the vector DM. Alternative conventions for this operator can be found in refs.~\cite{Dent:2015zpa,Catena:2019hzw,Gondolo:2020wge}. These choices do not affect the completeness of the operator basis to be introduced below.

\begin{table}[t]
    \renewcommand*{\arraystretch}{1.3}
	\tabcolsep 0.04in
    \centering
    \begin{tabular}{c|l|c|l}
    \toprule
    operator & definition & operator & definition \\\midrule
        $\mO_1$ & $\displaystyle \mathds{1}_{\DM}\mathds{1}_N$ & $\mO_{11}$ & $\displaystyle i\SDM\cdot\frac{\bq}{m_N}\mathds{1}_N$
        \\
        $\mO_3$ & $\displaystyle i\SN\cdot\left(\frac{\bq}{m_N}\times\vN\right)\mathds{1}_{\DM}$ & $\mO_{12}$ & $\displaystyle \SDM\cdot \left(\SN \times\vN \right)$
        \\
         $\mO_4$ & $\displaystyle \SDM\cdot \SN$ & $\mO_{13}$ & $\displaystyle i \left(\SDM\cdot \vN\right)\left(\SN\cdot \frac{\bq}{m_N}\right)$
        \\
         $\mO_5$ & $\displaystyle i\SDM\cdot\left(\frac{\bq}{m_N}\times\vN\right)\mathds{1}_N$ & $\mO_{14}$ & $\displaystyle i\left(\SDM\cdot \frac{\bq}{m_N}\right)\left(\SN\cdot \vN\right)$
        \\
         $\mO_6$ & $\displaystyle \left(\SDM\cdot\frac{\bq}{m_N}\right) \left(\SN\cdot\frac{\bq}{m_N}\right)$ &  $\mO_{15}$ & $\displaystyle -\left(\SDM\cdot \frac{\bq}{m_N}\right)\left[ \left(\SN\times \vN \right) \cdot \frac{\bq}{m_N}\right]$ 
        \\
         $\mO_7$ & $\displaystyle \SN\cdot \vN\mathds{1}_{\DM}$ & $\mO_{17}$ & $\displaystyle i \frac{\bq}{m_N} \cdot \boldsymbol{\mS} \cdot \vN \mathds{1}_N$
        \\
         $\mO_8$ & $\displaystyle \SDM\cdot \vN\mathds{1}_N$ & $\mO_{18}$ & $\displaystyle i \frac{\bq}{m_N} \cdot \boldsymbol{\mS}  \cdot \SN$
        \\
         $\mO_9$ & $\displaystyle i\SDM\cdot\left(\SN\times\frac{\bq}{m_N}\right)$ & $\mO_{19}$ & $\displaystyle \frac{\bq}{m_N} \cdot \boldsymbol{\mS} \cdot \frac{\bq}{m_N} \mathds{1}_N$
        \\
         $\mO_{10}$ & $\displaystyle i\SN\cdot\frac{\bq}{m_N}\mathds{1}_{\DM}$ & $\mO_{20}$ & $\displaystyle \left(\SN \times \frac{\bq}{m_N} \right) \cdot \boldsymbol{\mS} \cdot \frac{\bq}{m_N}$
        \\\bottomrule
    \end{tabular}
    \caption{NREFT operators up to first order in $\vN$ and second order in $\bq$, used throughout this work.}
    \label{tab:NREFT:operator}
\end{table}

Using the building blocks in eq.~\eqref{eq:building-blocks}, the complete operator basis up to first order in $\vN$ and second order in $\bq$ is given in table~\ref{tab:NREFT:operator}. For scalar and fermionic DM, we follow refs.~\cite{Fitzpatrick:2012ix,Anand:2013yka}, while for vector DM we adopt the operator basis introduced in refs.~\cite{Dent:2015zpa,Catena:2019hzw,Gondolo:2020wge,Liang:2025kkl}. The operator $\mathcal{O}_2$ is usually omitted because it does not arise from the leading-order non-relativistic reduction of a manifestly relativistic operator. Moreover, $\mathcal{O}_{16}$ is linearly dependent on $\mathcal{O}_{12}$ and $\mathcal{O}_{15}$~\cite{Anand:2013yka}. Therefore, the set of operators $\mathcal{O}_{1}, \ldots, \mathcal{O}_{20}$, excluding $\mathcal{O}_2$ and $\mathcal{O}_{16}$, is sufficient to describe the DM interactions induced by the DLEFT operators introduced in section~\ref{subsec:DLEFT}.

To obtain the Wilson coefficients of the NREFT operators, the DLEFT operators introduced in section~\ref{subsec:DLEFT} must be matched onto the NREFT basis. The DLEFT-NREFT matching for scalar and fermionic DM has been developed in refs.~\cite{Fitzpatrick:2012ix,Anand:2013yka}, while its extension to vector DM was studied in refs.~\cite{Dent:2015zpa,Catena:2019hzw,Gondolo:2020wge,Liang:2025kkl}. By translating the results of ref.~\cite{Liang:2025kkl} into our DLEFT basis, we obtain the following non-vanishing matching coefficients for the vector DM:
\begin{align}\label{eq:DLEFT-NREFT matching}
    c_1^N &= -\frac{1}{16m_X}F_S^{q/N}\,\left[4\left(L_{dX^2}+L_{dX^2}^*\right)-2i m^-\left(L_{dXX}^L+L_{dXX}^R\right)+m^+\left(L_{DdX^2}^L+L_{DdX^2}^R\right)\right] \nn\\
    &~\quad + \frac{1}{8m_X}A_{20}^{q/N}m_N\,\left(L_{DdX^2}^L+L_{DdX^2}^R\right) + \frac{1}{2m_X}F_G^N\,\mC_{GGXX}, \nn\\
    c_4^N &= -\frac{1}{2}G_A^{q/N}\,\left(L_{d\widetilde XX}^L+L_{d\widetilde XX}^R\right), \nn\\
    c_7^N &= -\frac{1}{3m_X}\tilde A_{20}^{q/N}m_N\,\left(L_{DdX^2}^R-L_{DdX^2}^L\right), \nn\\
    c_8^N &= -\frac{1}{4}F_1^{q/N}\,\left(L_{d\widetilde XX}^L-L_{d\widetilde XX}^R\right), \nn\\
    c_9^N &= \frac{1}{4}(F_1^{q/N}+F_2^{q/N})\,\left(L_{d\widetilde XX}^L-L_{d\widetilde XX}^R\right) - \frac{m_N^2}{4m_X^2} \tilde A_{20}^{q/N}\,\left(L_{DdX^2}^R-L_{DdX^2}^L\right), \nn\\
    c_{10}^N &= \frac{i}{16m_X}F_P^{q/N}\,\left[4\left(L_{dX^2}^*-L_{dX^2}\right)-2 m^+\left(L_{dXX}^L-L_{dXX}^R\right)+m^+\left(L_{DdX^2}^L-L_{DdX^2}^R\right)\right] \nn\\
    &~\quad- \frac{m_N}{6m_X}G_A^{q/N}\,\left(L_{dXX}^L-L_{dXX}^R\right) + \frac{1}{2m_X}F_{\widetilde G}^N\,\mC_{G\widetilde GXX}, \nn\\
    c_{17}^N &= \frac{m_N}{16m_X}F_1^{q/N}\,\left(L_{dXX}^L+L_{dXX}^R\right), \nn\\
    c_{18}^N &= \frac{m_N}{2m_X} G_A^{q/N}\,\left(L_{dXX}^L-L_{dXX}^R\right), \nn\\
    c_{20}^N &= \frac{m_N}{4m_X}(F_1^{q/N}+F_2^{q/N})\,\left(L_{dXX}^L+L_{dXX}^R\right).
\end{align}
Here, the quark-flavour indices have been suppressed with $m^\pm_{pr}\equiv m_p\pm m_r$, and $m^\pm L$ denotes $m^\pm_{pr}L_{pr}$ without summation over $p$ and $r$. The conventions for the nucleon matrix elements are taken from \texttt{DDCalc}~\cite{GAMBITDarkMatterWorkgroup:2017fax,GAMBIT:2018eea,FlavourPhysicsGroup:DDCalc}, except for the following four matrix elements, for which we adopt the convention of ref.~\cite{Liang:2025kkl}:
\begin{align}\label{eq:nucleon matrix element}
    \langle N(k')| ( \overline{q}\gamma_{(\mu} i \overleftrightarrow{D}_{\nu)} q) | N(k) \rangle & = \overline{u}_{N'} \left[A_{20}^{q/N}\, \gamma_{(\mu} K_{\nu)} + B_{20}^{q/N}\, \frac{i \sigma_{(\mu\rho} K_{\nu)} q^{\rho}}{2 m_N}+ C_{20}^{q/N}\, \frac{ 2 q_{(\mu}q_{\nu)}}{m_N} \right] u_{N},
    \nn\\
    \langle N(k')| ( \overline{q}\gamma_{(\mu}\gamma_5 i \overleftrightarrow{D}_{\nu)} q) | N(k) \rangle & = \overline{u}_{N'} \left[  \tilde A_{20}^{q/N}\, \gamma_{(\mu}\gamma_5 K_{\nu)} + \tilde B_{20}^{q/N}\, \frac{K_{(\mu}q_{\nu)}}{2 m_N}\gamma_5  \right] u_{N},
    \nn\\
    \langle N(k')| G^a_{\mu\nu}G^{\mu\nu,a} |N(k)\rangle&= F_G^{N}\, \overline{u}_{N'} u_N\,,
    \nn\\[0.2cm]
    \langle N(k')| G^a_{\mu\nu}\widetilde{G}^{\mu\nu,a}|N(k)\rangle&= F_{\widetilde G}^{N}\, \overline{u}_{N'} i \gamma_5 u_N\,,
\end{align}
where $q_\mu\equiv k'_\mu-k_\mu$ and $K_\mu\equiv k'_\mu+k_\mu$. We denote the symmetric traceless combination by $X_{(\mu} Y_{\nu)}\equiv \frac{1}{2}(X_\mu Y_\nu + X_\nu Y_\mu) - \frac{1}{4} g_{\mu\nu} X_\rho Y^\rho$. In the above expressions, the $q^2$ dependence of the nucleon form factors is always implied, e.g., $F_G^N \equiv F_G^N(q^2)$. For details of the non-relativistic reduction of the nucleon matrix elements, we refer the readers to refs.~\cite{Catena:2019hzw,DelNobile:2021wmp,Liang:2025kkl}.

\label{subsec:DM_nucleus_scattering}

In the non-relativistic limit, we can write the differential cross-section of the DM--nucleus scattering with respect to the recoil energy $E_R=\bq^2/(2m_T)$ as~\cite{DelNobile:2021wmp}
\begin{align}
    \frac{\text{d}\sigma_T}{\text{d}E_R}=\frac{1}{32\pi}\frac{1}{m_{\DM}^2 m_T}\frac{1}{v^2}\,\overline{|\mathcal M|^2},
\end{align}
where $m_T$ is the mass of the target nucleus and $v$ the speed of the incoming DM particle in the laboratory frame. The unpolarized amplitude squared factorizes as~\cite{DelNobile:2021wmp}
\begin{align}\label{eq:factorisation:new}
    \overline{|\mathcal M|^2}=\frac{m_T^2}{m_N^2}\sum_{N,N'=p,n}\sum_{X,Y} R_{XY}^{NN'}(\bq^2,\bv_T^{\perp2})\,F_{XY}^{NN'}(\bq^2),
\end{align}
where $F_{XY}^{NN'}=F_{YX}^{NN'}$ and $R_{XY}^{NN'}=R_{YX}^{NN'}$ denote the nuclear and DM response functions, respectively. Specific to our case, there are only eight non-vanishing nuclear response functions~\cite{Fitzpatrick:2012ix,Anand:2013yka,Catena:2015uha}: the six diagonal elements with the abbreviation $F_{X}^{NN'}\equiv F_{XX}^{NN'}$, where the subscript $X=M, \Sigma', \Sigma'', \tilde\Phi', \Phi'', \Delta$, while the other two off-diagonal interference terms are characterized with the subscripts $M\Phi''$ and $\Sigma'\Delta$. The corresponding DM response functions for scalar and fermionic DM were derived in refs.~\cite{Anand:2013yka,Catena:2015uua,Gondolo:2020wge,DelNobile:2021wmp}. For vector DM, we derive the response functions explicitly and our final expressions are consistent with the ones given in ref.~\cite{Liang:2025kkl}.\footnote{We note that a few minor typos appear in the expressions of ref.~\cite{Liang:2025kkl}. After correspondence with the authors, we confirm the corrected expressions, which are now in agreement with our results.} Introducing the abbreviation $R_{X}^{NN'}\equiv R_{XX}^{NN'}$, the DM response functions relevant for our analysis are given, respectively, as
\begin{align}\label{eq:DM response function}
    R_M^{NN'} &= c_1^N c_1^{N'} + \frac23\frac{\bq^2}{m_N^2}\vt^2 c_5^N c_5^{N'} + \frac23\vt^2 c_8^N c_8^{N'} + \frac23\frac{\bq^2}{m_N^2}c_{11}^N c_{11}^{N'}
    \nn\\
    &\quad+ \frac16\frac{\bq^2}{m_N^2}\vt^2 c_{17}^N c_{17}^{N'} + \frac29\frac{\bq^4}{m_N^4}c_{19}^N c_{19}^{N'}, \nn\\
    R_{\Sigma'}^{NN'} &= \frac16 c_4^N c_4^{N'} + \frac18\vt^2 c_7^N c_7^{N'} + \frac16\frac{\bq^2}{m_N^2}c_9^N c_9^{N'} + \frac1{12}\vt^2 c_{12}^N c_{12}^{N'} + \frac1{12}\frac{\bq^2}{m_N^2}\vt^2 c_{14}^N c_{14}^{N'} 
    \nn\\
    &\quad + \frac18\frac{\bq^2}{m_N^2}\vt^2 c_3^N c_3^{N'} - \frac1{12}\frac{\bq^2}{m_N^2}\vt^2\big(c_{12}^N c_{15}^{N'}+c_{15}^N c_{12}^{N'}\big) + \frac1{12}\frac{\bq^4}{m_N^4}\vt^2 c_{15}^N c_{15}^{N'} \nn \\ 
    &\quad + \frac1{24}\frac{\bq^2}{m_N^2}c_{18}^N c_{18}^{N'} + \frac1{24}\frac{\bq^4}{m_N^4}c_{20}^N c_{20}^{N'}, \nn\\[0.6em]
    R_{\Sigma''}^{NN'} &= \frac16 c_4^N c_4^{N'} + \frac16\frac{\bq^4}{m_N^4}c_6^N c_6^{N'} + \frac14\frac{\bq^2}{m_N^2}c_{10}^N c_{10}^{N'} + \frac16\vt^2 c_{12}^N c_{12}^{N'} + \frac16\frac{\bq^2}{m_N^2}\vt^2 c_{13}^N c_{13}^{N'} \nn\\
    &\quad + \frac1{18}\frac{\bq^2}{m_N^2}c_{18}^N c_{18}^{N'} + \frac16\frac{\bq^2}{m_N^2}\big(c_4^N c_6^{N'}+c_6^N c_4^{N'}\big), \nn\\
    R_{\tilde\Phi'}^{NN'} &= \frac16\frac{\bq^2}{m_N^2}c_{12}^N c_{12}^{N'} + \frac16\frac{\bq^4}{m_N^4}c_{13}^N c_{13}^{N'}, \nn\\
    R_{\Phi''}^{NN'} &= \frac14\frac{\bq^4}{m_N^4}c_3^N c_3^{N'} + \frac16\frac{\bq^2}{m_N^2}c_{12}^N c_{12}^{N'} + \frac16\frac{\bq^6}{m_N^6}c_{15}^N c_{15}^{N'} - \frac16\frac{\bq^4}{m_N^4}\big(c_{12}^N c_{15}^{N'}+c_{15}^N c_{12}^{N'}\big), \nn\\
    R_{\Delta}^{NN'} &= \frac23\frac{\bq^4}{m_N^4}c_5^N c_5^{N'} + \frac23\frac{\bq^2}{m_N^2}c_8^N c_8^{N'} + \frac16\frac{\bq^4}{m_N^4}c_{17}^N c_{17}^{N'}, \nn\\
    R_{M\Phi''}^{NN'} &= -\frac12\frac{\bq^2}{m_N^2}c_1^N c_3^{N'} - \frac13\frac{\bq^2}{m_N^2}c_{11}^N c_{12}^{N'} + \frac13\frac{\bq^4}{m_N^4}c_{11}^N c_{15}^{N'}, \nn\\
    R_{\Sigma'\Delta}^{NN'} &= -\frac13\frac{\bq^2}{m_N^2}c_4^N c_5^{N'} + \frac13\frac{\bq^2}{m_N^2}c_9^N c_8^{N'} - \frac1{12}\frac{\bq^4}{m_N^4}c_{20}^N c_{17}^{N'},
\end{align}
where, for a given nucleus target with $A$ nucleons, $\vt=1/A \sum_i v_i^\perp$ is simply the average of $\bv_N^\perp$ from all nucleons, and $\vt^2=\bv^2-\bq^2/(4\mu_{\DM T}^2)$ in the non-relativistic limit, with $\mu_{\DM T}$ being the reduced mass of the DM--nucleus system.

In this work, our numerical analyses of DM direct detection are based on the modified versions of \texttt{DDCalc}~\cite{GAMBITDarkMatterWorkgroup:2017fax,GAMBIT:2018eea,FlavourPhysicsGroup:DDCalc} and \texttt{DirectDM}~\cite{Bishara:2016hek,Bishara:2017pfq,Bishara:2017nnn,Brod:2017bsw,Brod:2018ust,Bishara:2018vix,FlavourPhysicsGroup:DirectDM}. In their original implementations, these packages support only scalar and fermionic DM. We have therefore extended \texttt{DirectDM} by implementing the vector-DM matching conditions in eq.~\eqref{eq:DLEFT-NREFT matching}, and modified \texttt{DDCalc} by incorporating the response functions for vector DM in eq.~\eqref{eq:DM response function}. In addition, using the numerical results of refs.~\cite{Blumlein:2010rn,Alexandrou:2020sml}, we have implemented in \texttt{DirectDM} the nucleon form factors $A_{20}^{q/N}$ and $\tilde{A}_{20}^{q/N}$ in eq.~\eqref{eq:nucleon matrix element}.

\section{Relic density}
\label{sec:Relic Density}

The observed DM relic density, $\Omega h^2=0.1200\pm 0.0012$~\cite{Planck:2018vyg}, provides a cosmological test of the top-flavoured DM interactions considered in this work. Assuming a thermal freeze-out, we calculate the relic abundance within the DLEFT and assess its consistency with this measurement. The relic abundance is determined by the thermally averaged cross-section for DM pair annihilation into SM particles, $\langle\sigma v\rangle$, which is given by~\cite{Gondolo:1990dk,Arcadi:2017kky}
\begin{align}\label{thermal-average-cross-section}
    \langle \sigma v\rangle=\frac{1}{8m_{\DM}^4TK^2_2(x)}\int^{\infty}_{4m_{\DM}^2} ds\,\sigma(s)\sqrt{s}\left(s-4m_{\DM}^2\right) K_1\left(\frac{\sqrt{s}}{T}\right),
\end{align}
where $x=m_{\DM}/T$, $K_1$ and $K_2$ are the modified Bessel functions of the second kind, and $s$ denotes the centre-of-mass energy squared of the DM pair. For the light DM masses considered in our phenomenological analysis, the energies relevant to annihilation during freeze-out are well below the EW scale. The annihilation cross-sections can be, therefore, evaluated within the DLEFT framework. Our calculation covers all 32 DLEFT operators with $Z_2$ symmetry introduced in section~\ref{subsec:DLEFT} for scalar, fermionic and vector DM. The results are provided electronically as supplementary files to the present article. In particular, the full dependence on $s$ is retained rather than expanded in the relative velocity $v$ of the annihilating DM particles~\cite{Gondolo:1990dk}. This is essential for the annihilation channel with final-state quarks being heavier than the DM, i.e., $m_q>m_{\DM}$ with $q=c,b$, for which a low-velocity expansion cannot reconstruct such a channel whose threshold lies above that expansion point and is therefore not applicable. With the thermally averaged annihilation cross-section at hand, the DM relic density can be calculated by solving the standard Boltzmann equation. We perform the numerical analysis using \texttt{micrOMEGAs}~\cite{Alguero:2023zol} and \texttt{DRAKE}~\cite{Binder:2021bmg}, together with the assistance of \texttt{DMAgent}~\cite{DMAgent:2026}. An independent cross-check using our own code yields results in good agreement with those obtained from these packages.

\begin{figure}
	\centering
	\includegraphics[width=0.99\linewidth]{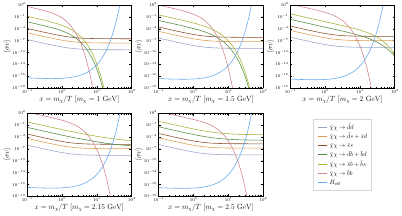}
	\caption{Thermally averaged cross-section of the annihilation process $\bar\chi \chi\to q_i q_j$ with $i,j=d,s,b$ and the effective Hubble rate $H_{\mathrm{eff}}(x)=H(x)/n_{\mathrm{eq}}$, as a function of the variable $x=m_\chi/T$, with five different DM masses. Their intersection point corresponds to the decoupling temperature of the process. }
    \label{fig:Cqchi_RD}
\end{figure}

To understand which annihilation channels control the relic abundance in the top-flavoured scenario, we examine their individual contributions to the thermally averaged cross-section. We take the DSMEFT operators $\mQ_{q\chi}$, $\mQ_{qXX}$, and $\mQ_{DqX^2}$ as three representative examples. After being matched onto the DLEFT at tree level, these operators induce DM pair annihilation into $d_i\bar d_j$ with $i,j=d,s,b$. The corresponding channel couplings are proportional to $V_{tj}V_{ti}^*$, as given in eq.~\eqref{eq:WC:DSMEFT operator}. The CKM factors entering the annihilation rates of DM pair into $b\bar b$, $s\bar b$, $d\bar b$, $s\bar s$, $d\bar s$ and $d\bar d$ therefore follow the hierarchy 
\begin{align} \label{eq:ckmhierarchy}
    |V_{tb}|^4:|V_{tb}V_{ts}^*|^2:|V_{tb}V_{td}^*|^2:|V_{ts}|^4:|V_{ts}V_{td}^*|^2:|V_{td}|^4 \sim 1:\lambda^4:\lambda^6:\lambda^8:\lambda^{10}:\lambda^{12},
\end{align} 
where $\lambda\approx0.23$ is the Wolfenstein parameter of the CKM matrix. The channels with the largest CKM factors also have the highest kinematic thresholds. We show in figure~\ref{fig:Cqchi_RD} the thermally averaged cross-sections $\langle\sigma v\rangle$ for the individual channels induced by $\mQ_{q\chi}$ and the effective Hubble rate $H_{\rm eff}(x)\equiv H(x)/n_{\rm eq}(x)$, as a function of the variable $x=m_\chi/T$, with five different DM masses. The intersection of the curve of each channel with $H_{\rm eff}$ provides an estimate of its decoupling temperature. The main features we observe are summarized as follows:
\begin{itemize}
    \item At high temperatures ($x\lesssim1$), all these channels are thermally accessible. Their annihilation rates follow the CKM hierarchy given by eq.~\eqref{eq:ckmhierarchy}, with their relative magnitudes governed primarily by the CKM factors and only weakly dependent on the DM mass.

    \item For the lower-mass benchmarks shown, which lie below the $s\bar b$ threshold, $m_\chi<(m_b+m_s)/2$, the channels containing a $b$ quark become Boltzmann suppressed for $x\gtrsim3$. Their estimated decoupling points, $x\simeq6\,$--$\,18$, occur well before those of the light-quark channels, $x\simeq20\,$--$\, 30$. The relic abundance is then controlled primarily by the $s\bar s$ channel, whose rate is suppressed by $|V_{ts}|^4$. For the benchmarks above this threshold, the $s\bar b$ and $d\bar b$ channels remain kinematically accessible at freeze-out, with their rates being two to three orders of magnitude larger than that of $s\bar s$. The $s\bar b$ channel is consequently the last to decouple. In contrast, the $b\bar b$ channel is kinematically forbidden in the zero-velocity limit for all the DM masses shown, since $m_\chi<m_b$.
\end{itemize}
The three operators $\mQ_{q\chi}$, $\mQ_{qXX}$, and $\mQ_{DqX^2}$ exhibit similar CKM hierarchies and threshold behaviours, but differ in their dependence on the energy $\sqrt{s}$. At high temperatures, the resulting thermally averaged cross-sections $\langle\sigma v\rangle$ scale approximately as $T^2$, $T^4$ and $T^6$, respectively. At late times, the annihilation into kinematically accessible channels is $s$-wave dominated for $\mQ_{q\chi}$ and $\mQ_{DqX^2}$, whereas it is $p$-wave dominated for $\mQ_{qXX}$, with $\langle\sigma v\rangle\propto1/x$.

\section{Indirect detection}
\label{sec:indirect detection}

In this section, we examine indirect constraints on the DM pair annihilation into SM particles from CMB observations~\cite{Boddy:2015efa} and gamma-ray line searches~\cite{Ackermann:2015lka}. For the CMB, the electromagnetic energy injection from DM annihilation can modify the ionization history of the Universe and hence the CMB anisotropies. For light DM, an approximate upper limit on the thermally averaged annihilation cross-section from Planck observations is quoted in ref.~\cite{Boddy:2015efa} as
\begin{align}\label{IDDM_sigmav}
    \langle \sigma v\rangle\lesssim  3\times 10^{-28} \left(\frac{m_{DM}}{\mathrm{GeV}}\right)\mathrm{cm^3/s},
\end{align}
where this approximate bound assumes that the $s$-wave annihilation of the DM pair exclusively into photons. For the gamma-ray line constraint, we use the Fermi-LAT $95\%$-CL limits on the diphoton annihilation cross-section from the R41 region for a standard Navarro-Frenk-White halo profile~\cite{Ackermann:2015lka}.

The quantity relevant to both constraints is the velocity-independent $s$-wave contribution to the thermally averaged diphoton annihilation cross-section defined by eq.~\eqref{thermal-average-cross-section}~\cite{Srednicki:1988ce,Gondolo:1990dk}. The relevant equations for the eight DLEFT operators that contribute directly to this process are listed in the last column of table~\ref{tab:DLEFT}. In our phenomenological analysis of the top-flavoured DSMEFT in section~\ref{sec:Numerical analysis}, we take the DSMEFT Wilson coefficients to be real. Under this assumption, the relevant DLEFT operators are $\mO_{FF\phi^2}$, $\mO_{FF\chi}$ and $\mO_{FFXX}$. At leading order in the relative-velocity expansion, the individual contributions to the annihilation cross-sections are given, respectively, by
\begin{align}
    \langle \sigma v_{\rm rel}\rangle\simeq\frac{8}{\pi}\frac{m_{\phi}^2}{v^4}|L_{FF\phi^2}|^2,\;\;
    \langle \sigma v_{\rm rel}\rangle \simeq\frac{1}{\pi}\frac{m_{\chi}^4}{v^6}|L_{FF\chi}|^2,\;\;
    \langle \sigma v_{\rm rel}\rangle\simeq\frac{8}{3\pi}\frac{m_{X}^2}{v^4}|L_{FFXX}|^2,
\end{align}
where $v$ denotes the EW vacuum expectation value.

\section{Final numerical analysis}
\label{sec:Numerical analysis}

In this section, we constrain the top-flavoured DM interactions within the EFT frameworks described above using the flavour observables of section~\ref{sec:flavour}, the constraints from DM direct detection discussed in section~\ref{sec:direct detection}, the relic density of section~\ref{sec:Relic Density}, and the CMB and Fermi-LAT bound of section~\ref{sec:indirect detection}. For the input parameters relevant to flavour observables, we refer the readers to ref.~\cite{Hou:2024vyw} and the references therein, while for those relevant to DM direct detection, they are mainly taken from ref.~\cite{Liang:2025kkl}. For convenience, we collect in appendix~\ref{app:hadronic matrix element} the relevant definitions and conventions used throughout this work, and the numerical values of the main input parameters are listed in table~\ref{tab:input}. Throughout the analysis, we fix $\Lambda=1\TeV$ and consider one DSMEFT operator at a time, with a real Wilson coefficient. We focus on the eleven operators in table~\ref{tab:DSMEFT} that contain two identical DM fields, namely $\mQ_{u\phi^2}$, $\mQ_{q\chi}$, $\mQ_{u\chi}$, $\mQ_{u\chi^2}$, $\mQ_{uX^2}$, $\mQ_{qXX}$, $\mQ_{uXX}$, $\mQ_{q\widetilde XX}$, $\mQ_{u\widetilde XX}$, $\mQ_{DqX^2}$ and $\mQ_{DuX^2}$, for which the DM is stable. Operators containing a single DM field or two distinct DM states ($\mQ_{\phi q}$, $\mQ_{\phi u}$, $\mQ_{uHX^2}$) are constrained only by the flavour observables discussed in section~\ref{sec:flavour}. We scan the DM masses up to $2.5\GeV$, covering the kinematically allowed range for the DM pair production in $B\to K+\inv$ decays.

\begin{figure}[t]
    \centering
    \includegraphics[width=0.96\linewidth]{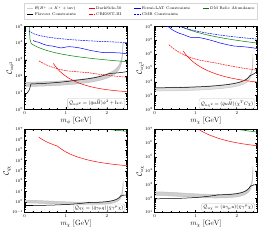}
    \caption{Combined constraints on the Wilson coefficients of the DSMEFT operators involving scalar (upper left) and fermionic (other panels) DM, as a function of the DM mass, for $\Lambda=1\TeV$. The gray band is the $2\sigma$ region favoured by the Belle II measurement of $\mB(B^+\to K^++\inv)$, and the black line is the upper limit from the other flavour observables in table~\ref{tab:observables}. The red and blue lines are the upper limits from DM direct detection (CRESST-III and DarkSide-50) and from the CMB and Fermi-LAT bound, respectively. The green line is the contour required for reproducing the observed relic density $\Omega h^2=0.12$; for $\mQ_{q\chi}$ it lies above the plotted range.}
    \label{fig:total:SF}
\end{figure}

\begin{figure}[t]
    \centering
    \includegraphics[width=0.99\linewidth]{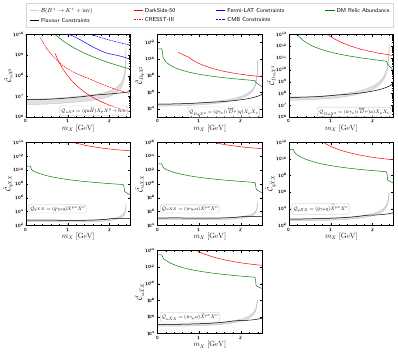}
    \caption{Same as in figure~\ref{fig:total:SF}, but for the DSMEFT operators involving vector DM, in terms of the rescaled Wilson coefficients $\tilde\mC_i$ defined in eq.~\eqref{eq:rescaledWCs}. The CMB and Fermi-LAT bound is shown only for $\mQ_{uX^2}$, the only vector-DM operator that induces $XX\to\gamma\gamma$ at leading order in the relative velocity expansion.}
    \label{fig:total:V}
\end{figure}

Our main results are shown in figures~\ref{fig:total:SF} and \ref{fig:total:V}. In each panel, the gray band is the $2\sigma$ region required to accommodate the Belle II measurement of $\mB(B^+\to K^++\inv)$, and the black line is the upper limit from the other flavour observables in table~\ref{tab:observables}. The red line shows the $90\%$~CL upper bound from direct detection experiments, set by CRESST-III~\cite{CRESST:2019jnq} for $0.35\GeV\lesssim m_{\DM}\lesssim0.8\GeV$ and by DarkSide-50~\cite{DarkSide:2018bpj} for $m_{\DM}>0.8\GeV$, the transition being visible as a kink at $m_{\DM}\simeq0.8\GeV$. The blue line is the CMB limit from eq.~\eqref{IDDM_sigmav}, and the green line is the contour required to reproduce the observed relic density $\Omega h^2=0.12$~\cite{Planck:2018vyg}: below it DM is overproduced, while above it DM is only a fraction of the total relic abundance.

In DM direct detection, the DM--nucleon scattering receives contributions from flavour-diagonal light-quark operators and gluon operators. The light-quark interactions can arise at tree level with CKM suppression or be loop-induced, while the gluon interactions are generated at one-loop level. We find that the strength of the constraints from DM direct detection depends strongly on the Lorentz structure of the underlying DSMEFT operator:
\begin{itemize}
    \item For the scalar-type operators $\mQ_{u\phi^2}$, $\mQ_{u\chi^2}$ and $\mQ_{uX^2}$, the nucleon couplings arise from the top-loop induced gluon operators and from Higgs-mediated light-quark scalar currents. Neither is CKM suppressed, and both benefit from the large scalar nucleon form factors. Direct detection is therefore competitive with the rare $B$-meson decays and, as can be seen from figures~\ref{fig:total:SF} and \ref{fig:total:V}, excludes the entire Belle~II band for $m_\phi\gtrsim1.35\GeV$, $m_\chi\gtrsim2.3\GeV$ and $m_X\gtrsim1.8\GeV$, respectively. Since the constraints from direct detection become rapidly stronger as the DM mass increases while the flavour bounds become flat or weak, the two probes are complementary. For example, for $\mQ_{u\phi^2}$, the constraints from $K\to\pi+\inv$ decays and direct detection impose lower and upper bounds on the DM mass, respectively; this restricts the Belle II-favoured range to $0.2\GeV\lesssim m_\phi\lesssim1.35\GeV$.

    \item The DSMEFT operators $\mQ_{q\chi}$ and $\mQ_{u\chi}$ induce the DLEFT vector-current operators $\mO_{d\chi}^{V,LR}$, $\mO_{d\chi}^{V,RR}$ and their up-type counterparts $\mO_{u\chi}^{V,LR}$, $\mO_{u\chi}^{V,RR}$ through tree-level and one-loop matching. The corresponding flavour-diagonal Wilson coefficients for light-quark operators are CKM- or loop-suppressed. Specifically, $\mQ_{q\chi}$ generates a tree-level contribution to the $d$-quark Wilson coefficient proportional to $|V_{td}|^2\simeq7\times10^{-5}$, whereas its contributions to the $u$-quark Wilson coefficients arise only at the one-loop level. For $\mQ_{u\chi}$, the Wilson coefficients of both the $u$- and $d$-quark operators are loop-suppressed, as are those governing the $b\to s$ transition. Numerically, the upper bounds on the Wilson coefficients from direct detection lie four to five orders of magnitude above the Belle~II band and do not provide further constraint on this region. The same holds for the derivative operators $\mQ_{DqX^2}$ and $\mQ_{DuX^2}$, whose flavour-diagonal matrix elements are proportional to the twist-2 nucleon form factors $A_{20}^{q/N}$, $\tilde A_{20}^{q/N}$ and to the light-quark masses, with the resulting limits being three to five orders of magnitude above the band. The remaining four vector-DM operators, $\mQ_{qXX}$, $\mQ_{uXX}$, $\mQ_{q\widetilde XX}$ and $\mQ_{u\widetilde XX}$, match only onto the velocity-suppressed NREFT operators; consequently, their direct-detection bounds lie eight orders of magnitude above the Belle~II band. For all eight operators, the Belle II-favoured region is therefore constrained by flavour observables alone.
\end{itemize}
Among the direct detection experiments for light DM considered here, the DarkSide-50 provides the stronger bound for $m_{\DM}\gtrsim0.8\GeV$, while the CRESST-III supplies the constraints at the lower masses probed here. Future experiments with lower detection thresholds could extend this reach and further test the light-DM parameter space accessible through $B\to K+\inv$ decays.

For indirect detection, the relevant $s$-wave annihilation channel is $\DM\,\DM\to\gamma\gamma$. With real Wilson coefficients and at the order considered here, this channel is induced only by the DSMEFT operators $\mQ_{u\phi^2}$, $\mQ_{u\chi^2}$ and $\mQ_{uX^2}$ through the top-loop matching contributions in eqs.~\eqref{eq:WC:S:gmgm}--\eqref{eq:WC:V:gmgm}. The CMB and Fermi-LAT bound derived from eq.~\eqref{IDDM_sigmav} becomes weaker at lower DM masses, as shown by the blue lines in figures~\ref{fig:total:SF} and \ref{fig:total:V}. These upper bounds lie two to three orders of magnitude above the regions favoured by Belle~II and the corresponding upper bounds from direct detection. Therefore, the CMB and Fermi-LAT constraints considered here do not exclude the parameter space favoured by the Belle~II excess.

Turning to the relic density discussed in section~\ref{sec:Relic Density}, we show the parameter values that reproduce the observed abundance as green contours in figures~\ref{fig:total:SF} and \ref{fig:total:V}. These contours lie above the Belle~II region over essentially the whole DM mass range. Within the region favoured by Belle~II, the standard thermal freeze-out through the top-flavoured interaction alone therefore overproduces the DM states. The only exception is the upper edge of the kinematically allowed range, $m_{\DM}\simeq2.3$--$2.4\GeV$, where the opening of the $s\bar b$ and $d\bar b$ channels at freeze-out makes the contour drop by more than an order of magnitude and cross the Belle~II band (e.g., for $\mQ_{DqX^2}$ and $\mQ_{DuX^2}$), so that the observed relic abundance and the $B^+\to K^++\inv$ excess can be reproduced simultaneously in a narrow mass window.\footnote{For $\mQ_{u\phi^2}$ and $\mQ_{uX^2}$, the relic-density contours also cross the Belle~II band in this narrow high-mass range. In these cases, however, the annihilation is dominated by the loop-induced $\gamma\gamma$ and $gg$ final states, and the contours vary smoothly over the entire DM mass range rather than dropping at the $s\bar b$ and $d\bar b$ thresholds.} Elsewhere, a top-flavoured explanation of the Belle~II excess requires additional annihilation channels, e.g., into dark-sector states.

\section{Conclusion}
\label{sec:conclusion}

In this work, we have investigated the light top-flavoured DM within the DSMEFT framework, considering scalar, fermionic and vector DM, together with ALPs. Their interactions with the SM particles are specified through third-generation quark operators at the EW scale, while the light invisible particles remain dynamical below this scale. This setup correlates rare $B$- and $K$-meson decays with missing energy, direct and indirect detection, and the thermal relic abundance through a common set of DSMEFT Wilson coefficients. To describe these observables within such a single framework, we have constructed the EFT chain of DSMEFT $\to$ DLEFT $\to$ NREFT, taking into account the proper matching and running effects across different scales.

The one-loop matching of the DSMEFT onto the DLEFT is performed using the standard off-shell matching procedure. The resulting DLEFT Wilson coefficients, collected in appendix~\ref{app:matching coefficients}, are used to describe the flavour-changing and flavour-diagonal quark interactions, as well as the DM couplings to gluons and photons. After QCD evolution to the appropriate low-energy scales, these results provide a consistent basis for our phenomenological analysis of rare $B$- and $K$-meson decays with missing energy, direct and indirect detection, and the DM relic density. For direct detection, we have further matched the DLEFT onto the NREFT, including the spin-1-specific operators, and derived the correct response functions required for vector DM. These results have been implemented into the publicly available extended versions of \texttt{DirectDM}~\cite{FlavourPhysicsGroup:DirectDM} and \texttt{DDCalc}~\cite{FlavourPhysicsGroup:DDCalc}. We have also calculated the annihilation cross-sections induced by all 32 $Z_2$-symmetric DLEFT operators considered here, retaining their full energy dependence to account for thermal access to final states above the zero-velocity threshold.

For such a top-flavoured DM scenario, a generic feature of the DSMEFT-to-DLEFT matching is that, for each DSMEFT operator, all resulting down-type FCNC interactions are controlled by a single coupling, $[\mC_i]_{33}$, and inherit the CKM structure of the SM top-quark loops, $V_{tj}V_{ti}^{*}$. Contributions involving scalar quark currents are additionally suppressed by the external down-type quark masses. As a result, our numerical analysis with $\Lambda=1\TeV$ found that only operators contributing to the relevant transitions at tree level can accommodate the Belle~II excess with coefficients of $\mathcal{O}(1)$. The loop-induced operators instead require coefficients of $\mathcal{O}(10$--$100)$, pushing the EFT to the boundary of its validity and challenging a weakly coupled interpretation at this scale. We have also investigated constraints from other flavour observables on the parameter regions favoured by the Belle~II excess. Among the relevant processes, the decay $B^0\to K^{*0}+\inv$ provides the leading constraint over large portions of the parameter space. For some operators, the constraint from $K^+\to\pi^++\inv$ dominates at low DM masses, whereas the most sensitive channels at large DM masses are from $B_s\to\inv$, $B^+\to\pi^++\inv$, and $\Upsilon(1S)\to\inv$.

We have also studied constraints on the DSMEFT operators from the direct and indirect detection, as well as the DM relic density. It is found that direct detection supplies complementary sensitivity, particularly for operators containing scalar quark currents. These operators generate gluonic interactions without CKM suppression, making the CRESST-III and DarkSide-50 bounds competitive with the rare $B$- and $K$-meson decays with missing energy. For $\mQ_{u\phi^2}$, $\mQ_{u\chi^2}$ and $\mQ_{uX^2}$, the direct-detection limits exclude the entire Belle~II-favoured band for the DM mass above approximately $1.35$, $2.3$, and $1.8\GeV$, respectively. For the other operators examined, CKM factors, loop suppression, or velocity-dependent scattering substantially weaken the direct-detection sensitivity. Under the assumption of real Wilson coefficients, the CMB and Fermi-LAT diphoton-annihilation bounds considered here do not exclude the Belle~II-favoured regions for any DSMEFT operator considered. Regarding the DM relic density, the annihilation into light SM degrees of freedom is suppressed by CKM or loop factors. As a result, throughout most of the Belle~II-favoured parameter space, annihilation through the top-flavoured interactions alone is too weak to produce the observed abundance by standard thermal freeze-out. For the selected operators, the relic-density contours intersect the Belle~II band in a narrow region near $m_{\DM}\simeq 2.3$–$2.4\GeV$. Away from such regions, reproducing both the Belle~II excess and the observed thermal abundance requires additional annihilation mechanisms. In an ultraviolet-complete model, the DM may also interact with other SM particles or dark-sector states, or its coupling to the top quark may arise through a light mediator. Such additional interactions or light degrees of freedom may open efficient annihilation channels and thus reconcile the top-flavoured explanation of the Belle~II excess with the observed relic abundance.

Our analysis can be extended in several directions. We have fixed the DSMEFT Wilson coefficients at the EW scale; the RG evolution from the NP scale down to the EW scale, which would mix the top-quark operators into operators containing Higgs and gauge bosons, can be included in a more complete analysis. This would, in turn, require a complete DSMEFT-to-DLEFT matching including all operators. Although the particles mediating the DM--SM interactions are assumed to be heavy in the present EFT framework, many of our results can be adapted to scenarios with light mediators. Studying these scenarios would further clarify how flavour observables probe light DM in concert with direct and indirect detection and cosmological measurements of the DM relic density.

\acknowledgments
We thank Qing-Hong Cao, Hao-Lin Li, Jia Liu, Xiao-Dong Ma and Jiang-Hao Yu for useful discussions. This work is supported by the National Natural Science Foundation of China under Grant Nos.~12635005, 12475094, 12135006, 12575099, and 12505109, the Fundamental Research Funds for the Central Universities under Grand No.~XJ2026000201, as well as the Science and Technology Innovation Leading Talent Support Program of Henan Province under Grant No.~254000510039.

\appendix

\section{Matching coefficients}
\label{app:matching coefficients}

In this appendix, we summarize the matching coefficients obtained in section~\ref{sec:matching}. For each DLEFT operator, the total contributions to its Wilson coefficient is given by
\begin{align}
    [L_i]_{pr} + \delta_{pr} [L_i]_{pp} \qquad (\text{no summation over } p).
\end{align}

\boxtitle{f\text{-}f\text{-}\phi}
\begin{flalign}\label{eq:WC:S:1}
    \left[L_{d\phi}\right]_{pr}&=\frac{V_{tr}V_{tp}^*}{16\sqrt{2}\pi^2}\frac{m_r m_t}{v \Lambda}\big(  \left[\mC_{u\phi}\right]_{33} S_1+\left[\mC_{u\phi}\right]_{33}^*  S_2  \big).&
\end{flalign}

\boxtitle{f\text{-}f\text{-}\phi\text{-}\phi}
\begin{flalign}
    \left[L_{d\phi^2}\right]_{pr}=&\frac{V_{tr}V_{tp}^*}{16\sqrt{2}\pi^2}\frac{m_r m_t}{ \Lambda^2}\big( \left[\mC_{u\phi^2}\right]_{33} S_1 + \left[\mC_{u\phi^2}\right]_{33}^* S_2 \big),&\label{eq:WC:dphi2:pr}\\
    \left[L_{\phi d}^R\right]_{pr}=&0,&\label{eq:WC:phid:R:pr}\\
    \left[L_{\phi d}^L\right]_{pr}=&V_{tr}V_{tp}^* \frac{v^2}{\Lambda^2}\left[\mC_{\phi q}\right]_{33}+\frac{V_{tr}V_{tp}^*}{16\pi^2}\frac{m_t^2}{\Lambda^2} \left[\mC_{\phi u}\right]_{33} S_3,&\label{eq:WC:S2:pr}
    \\[1em]
    \left[L_{d\phi^2}\right]_{pp}=&\frac{3}{4\sqrt{2} \pi^2}\frac{m_p m_t^3}{m_h^2 \Lambda^2} \Re\left[\mC_{u\phi^2}\right]_{33}S_4-\frac{i}{8\sqrt{2} \pi^2}\frac{m_p m_t}{\Lambda^2} \Im\left[\mC_{u\phi^2}\right]_{33} S_5,&\label{eq:WC:S:2:pp}\\
     \left[L_{u\phi^2}\right]_{pp}=&\frac{3}{4\sqrt{2} \pi^2}\frac{m_p m_t^3}{m_h^2 \Lambda^2}\Re\left[\mC_{u\phi^2}\right]_{33} S_4+\frac{i}{8\sqrt{2} \pi^2}\frac{m_p m_t}{\Lambda^2}\Im\left[\mC_{u\phi^2}\right]_{33}S_5.&\label{eq:WC:uphi2:pp}
\end{flalign}

\boxtitle{f\text{-}f\text{-}a}
\begin{flalign}
    \left[L_{da}^{L}\right]_{pr}=&V_{tr}V_{tp}^*\left[\mC_{qa}\right]_{33}
    +\frac{V_{tr}V_{tp}^*}{16\pi^2} \frac{m_t^2}{v^2}\left[\mC_{ua}\right]_{33} S_3,&\label{eq:WC:ALP:L:pr}\\
    \left[L_{da}^{R}\right]_{pr}=&0.&\label{eq:WC:ALP:R:pr}
\end{flalign}

It is noted that the DLEFT operator $\mO_{\phi d}^L$ receives contribution from the DSMEFT operator $\mQ_{\phi q}$ at the tree-level, but from $\mQ_{\phi u}$ at the one-loop level. \\

\boxtitle{f\text{-}f\text{-}\chi\text{-}\chi}
\begin{flalign}
     \left[L_{d\chi}^{S,RR}\right]_{pr}&=\frac{V_{tr}V_{tp}^*}{16\sqrt{2}\pi^2}\frac{m_r m_t v}{\Lambda^3}\left[\mC_{u\chi^2}\right]_{33} S_1,&\label{eq:WC:dchi:SRR:pr}\\
	\left[L_{d\chi}^{S,LR}\right]_{pr}&=\frac{V_{tr}V_{tp}^*}{16\sqrt{2}\pi^2}\frac{m_p m_t v}{\Lambda^3}\left[\mC_{u\chi^2}\right]_{33} S_2,&\label{eq:WC:dchi:SLR:pr}\\
    \left[L_{d\chi}^{V,LR}\right]_{pr}&=V_{tr}V_{tp}^*\frac{v^2}{\Lambda^2}\left[\mC_{q\chi}\right]_{33}+\frac{V_{tr}V_{tp}^*}{16\pi^2}\frac{m_t^2}{\Lambda^2} \left[\mC_{u\chi}\right]_{33} S_3,&\label{eq:WC:dchi:VLR:pr}\\
    \left[L_{d\chi}^{V,RR}\right]_{pr}&=0,&
    \nn\\
    \left[L_{d\chi}^{T,RR}\right]_{pr}&=0,& \label{eq:WC:F:pr}
    \\[1em]
    \left[L_{d\chi}^{V,RR}\right]_{pp}&=\frac{s_W^2}{27\pi^2}\frac{m_W^2}{\Lambda^2}\left[\mC_{u\chi}\right]_{33} S_5+\frac{s_W^2}{12\pi^2}\frac{m_t^2}{\Lambda^2}\left[\mC_{u\chi}\right]_{33} S_5,&\label{eq:WC:dchi:VRR:pp}\\
    \left[L_{d\chi}^{V,LR}\right]_{pp}&=\frac{s_W^2}{27\pi^2}\frac{m_W^2}{\Lambda^2}\left[\mC_{u\chi}\right]_{33} S_5+\frac{2s_W^2-3}{24\pi^2}\frac{m_t^2}{\Lambda^2}\left[\mC_{u\chi}\right]_{33} S_5,&\label{eq:WC:dchi:VLR:pp}\\
    \left[L_{u\chi}^{V,RR}\right]_{pp}&=-\frac{2s_W^2}{27\pi^2}\frac{m_W^2}{\Lambda^2}\left(\left[\mC_{u\chi}\right]_{33}+\left[\mC_{q\chi}\right]_{33}\right) S_5-\frac{s_W^2}{6\pi^2}\frac{m_t^2}{\Lambda^2}\left(\left[\mC_{u\chi}\right]_{33}-\left[\mC_{q\chi}\right]_{33}\right) S_5,&\label{eq:WC:uchi:VRR:pp}\\
    \left[L_{u\chi}^{V,LR}\right]_{pp}&=-\frac{2s_W^2}{27\pi^2}\frac{m_W^2}{\Lambda^2}\left(\left[\mC_{u\chi}\right]_{33}+\left[\mC_{q\chi}\right]_{33}\right) S_5+\frac{3-4s_W^2}{24\pi^2}\frac{m_t^2}{\Lambda^2}\left(\left[\mC_{u\chi}\right]_{33}-\left[\mC_{q\chi}\right]_{33}\right) S_5,&\label{eq:WC:uchi:VLR:pp}\\
    \left[L_{d\chi}^{S,LR}\right]_{pp}&=-\frac{3}{8\sqrt{2}\pi^2}\frac{m_p m_t^3 v}{m_h^2 \Lambda^3} \left[\mC_{u\chi^2}\right]_{33} S_4-\frac{1}{16\sqrt{2} \pi^2}\frac{m_p m_t v}{\Lambda^3} \left[\mC_{u\chi^2}\right]_{33} S_5,&\label{eq:WC:dchi:SLR:pp}\\
    \left[L_{d\chi}^{S,RR}\right]_{pp}&=-\frac{3}{8\sqrt{2}\pi^2}\frac{m_p m_t^3 v}{m_h^2 \Lambda^3} \left[\mC_{u\chi^2}\right]_{33} S_4+\frac{1}{16\sqrt{2} \pi^2}\frac{m_p m_t v}{\Lambda^3} \left[\mC_{u\chi^2}\right]_{33} S_5,&\label{eq:WC:dchi:SRR:pp}\\
    \left[L_{u\chi}^{S,LR}\right]_{pp}&=-\frac{3}{8\sqrt{2}\pi^2}\frac{m_p m_t^3 v}{m_h^2 \Lambda^3} \left[\mC_{u\chi^2}\right]_{33} S_4+\frac{1}{16\sqrt{2} \pi^2}\frac{m_p m_t v}{\Lambda^3} \left[\mC_{u\chi^2}\right]_{33} S_5,&\label{eq:WC:uchi:SLR:pp}\\
    \left[L_{u\chi}^{S,RR}\right]_{pp}&=-\frac{3}{8\sqrt{2}\pi^2}\frac{m_p m_t^3 v}{m_h^2 \Lambda^3} \left[\mC_{u\chi^2}\right]_{33} S_4-\frac{1}{16\sqrt{2} \pi^2}\frac{m_p m_t v}{\Lambda^3} \left[\mC_{u\chi^2}\right]_{33} S_5.&\label{eq:WC:uchi:SRR:pp}
\end{flalign}

\boxtitle{f\text{-}f\text{-}X}
\begin{flalign}
    \left[L_{dX}^T\right]_{pr}=
    &+\frac{V_{tr}V_{tp}^*}{32\pi^2}\frac{m_p}{v}\left( \left[\mC_{uX}\right]_{33} V_3+\left[\mC_{qX}\right]_{33} V_4\right)&
    \nn\\
    &+\frac{V_{tr}V_{tp}^*}{16\sqrt{2}\pi^2}\frac{m_p m_t}{\Lambda^2}\left( \left[\mC_{uHX}\right]_{33} V_1 + \left[\mC_{uHX}\right]_{33}^* V_2 \right)&
    \nn\\
    &+\frac{V_{tr}V_{tp}^*}{64\pi^2}\frac{m_p v}{\Lambda^2}\left(\left[\mC_{HuX}\right]_{33} V_3 +\left[\mC_{HqX}^{(1)}\right]_{33} V_4 - \left[\mC_{HqX}^{(3)}\right]_{33} V_4 \right),&\label{eq:WC:dX:T:pr}\\
    \left[L_{dX}^{L}\right]_{pr}=&V_{tr}V_{tp}^*\left(\left[\mC_{qX}\right]_{33}+\frac{v^2}{2\Lambda^2}\left[\mC_{HqX}^{(1)}\right]_{33}-\frac{v^2}{2\Lambda^2}\left[\mC_{HqX}^{(3)}\right]_{33}\right)&
    \nn\\
    &+\frac{V_{tr}V_{tp}^*}{16\pi^2} \left(\frac{m_t^2}{v^2}\left[\mC_{uX}\right]_{33}+\frac{m_t^2}{2\Lambda^2}\left[\mC_{HuX}\right]_{33}\right) S_3,&\label{eq:WC:dX:L:pr}\\
    \left[L_{dX}^{R}\right]_{pr}=&0.&\label{eq:WC:dX:R:pr}
\end{flalign}

\boxtitle{f\text{-}f\text{-}X\text{-}X}
\begin{flalign}
    \left[L_{dX^2}\right]_{pr}=&\frac{V_{tr}V_{tp}^*}{16\sqrt{2}\pi^2}\frac{m_r m_W}{\Lambda^2}\left(\left[\mC_{uX^2}\right]_{33} S_1 + \left[\mC_{uX^2}\right]_{33}^* S_2\right)\nn\\
     &+\frac{V_{tr}V_{tp}^*}{24\pi^2}\frac{m_r}{v \Lambda^2}\left(m_t^2\Re\left[\mC_{DuX^2}\right]_{33} V_5+ m_W^2\Re\left[\mC_{DqX^2}\right]_{33}V_6\right),&\label{eq:WC:dX2:pr}\\
    \left[L_{dX^2}^T\right]_{pr}=&\frac{V_{tr}V_{tp}^*}{16\sqrt{2}\pi^2}\frac{m_r m_t}{\Lambda^2}\left(\left[\mC_{uHX^2}\right]_{33} V_{10} +\left[\mC_{uHX^2}\right]_{33}^* V_9\right),&\label{eq:WC:dX2:T:pr}\\
    \left[L_{dXX}^L\right]_{pr}=&V_{tr}V_{tp}^*\frac{v^2}{\Lambda^2}\left[\mC_{qXX}\right]_{33}+\frac{V_{tr}V_{tp}^*}{16\pi^2}\frac{m_t^2}{\Lambda^2}\left[\mC_{uXX}\right]_{33} S_3 ,&\label{eq:WC:dXX:L:pr}\\
    \left[L_{dXX}^R\right]_{pr}= & 0,&
    \nn\\
    \left[L_{d \widetilde{X} X}^L\right]_{pr}= & V_{tr}V_{tp}^*\frac{v^2}{\Lambda^2}\left[\mC_{q\widetilde{X}X}\right]_{33}+\frac{V_{tr}V_{tp}^*}{16\pi^2}\frac{m_t^2}{\Lambda^2}\left[\mC_{u\widetilde{X}X}\right]_{33} S_3,&\label{eq:WC:dXtX:L:pr}\\
    \left[L_{d \widetilde{X} X}^R\right]_{pr}= & 0,&\label{eq:WC:dXtX:R:pr}\\
    \left[L_{DdX^2}^L\right]_{pr}=&2V_{tr}V_{tp}^*\frac{v^2}{\Lambda^2}\Re\left[\mC_{DqX^2}\right]_{33}+\frac{V_{tr}V_{tp}^*}{48\pi^2}\frac{m_t^2}{\Lambda^2}\left(\left[\mC_{DuX^2}\right]_{33} V_7 + \left[\mC_{DuX^2}\right]_{33}^* V_8\right),&\label{eq:WC:DdX2:L:pr}\\
    \left[L_{DdX^2}^R\right]_{pr}=&0,&\label{eq:WC:V:2:pr}
    \\[1em]
    \left[L_{dX^2}\right]_{pp}=&\frac{3}{4\sqrt{2} \pi^2}\frac{m_p m_t^3}{m_h^2 \Lambda^2}\Re\left[\mC_{uX^2}\right]_{33}S_4-\frac{i}{8\sqrt{2} \pi^2}\frac{m_p m_t}{\Lambda^2}\Im\left[\mC_{uX^2}\right]_{33} S_5,&\label{eq:WC:dX2:pp}\\
    \left[L_{dXX}^L\right]_{pp} = & -\frac{s_W^2}{27\pi^2}\frac{m_W^2}{\Lambda^2}\left[\mC_{uXX}\right]_{33} S_5-\frac{2s_W^2-3}{24\pi^2}\frac{m_t^2}{\Lambda^2}\left[\mC_{uXX}\right]_{33} S_5,&\label{eq:WC:dXX:L:pp}\\
    \left[L_{dXX}^R\right]_{pp} = & -\frac{s_W^2}{27\pi^2}\frac{m_W^2}{\Lambda^2}\left[\mC_{uXX}\right]_{33} S_5-\frac{s_W^2}{12\pi^2}\frac{m_t^2}{\Lambda^2}\left[\mC_{uXX}\right]_{33} S_5,&\label{eq:WC:dXX:R:pp}\\
    \left[L_{DdX^2}^R\right]_{pp} = &-i\frac{s_W^2}{27\pi^2}\frac{m_W^2}{\Lambda^2}\Im\left[\mC_{DuX^2}\right]_{33} S_5-i\frac{s_W^2}{12\pi^2}\frac{m_t^2}{\Lambda^2}\Im\left[\mC_{DuX^2}\right]_{33} S_5,&\label{eq:WC:DdX2:R:pp}\\
    \left[L_{DdX^2}^L\right]_{pp} = &-i\frac{s_W^2}{27\pi^2}\frac{m_W^2}{\Lambda^2}\Im\left[\mC_{DuX^2}\right]_{33} S_5-i\frac{2s_W^2-3}{24\pi^2}\frac{m_t^2}{\Lambda^2}\Im\left[\mC_{DuX^2}\right]_{33} S_5,&\label{eq:WC:DdX2:L:pp}\\
    \left[L_{uX^2}\right]_{pp}=&\frac{3}{4\sqrt{2} \pi^2}\frac{m_p m_t^3}{m_h^2 \Lambda^2}\Re\left[\mC_{uX^2}\right]_{33}S_4+\frac{i}{8\sqrt{2} \pi^2}\frac{m_p m_t}{\Lambda^2}\Im\left[\mC_{uX^2}\right]_{33} S_5,&
    \nn\\
    \left[L_{uXX}^L\right]_{pp} = & \frac{2s_W^2}{27\pi^2}\frac{m_W^2}{\Lambda^2}\left[\mC_{uXX}\right]_{33} S_5+\frac{4s_W^2-3}{24\pi^2}\frac{m_t^2}{\Lambda^2}\left[\mC_{uXX}\right]_{33} S_5,&
    \nn\\
    \left[L_{uXX}^R\right]_{pp} = & \frac{2s_W^2}{27\pi^2}\frac{m_W^2}{\Lambda^2}\left[\mC_{uXX}\right]_{33} S_5+\frac{s_W^2}{6\pi^2}\frac{m_t^2}{\Lambda^2}\left[\mC_{uXX}\right]_{33} S_5,&
    \nn\\
    \left[L_{DuX^2}^R\right]_{pp} = &i\frac{2s_W^2}{27\pi^2}\frac{m_W^2}{\Lambda^2}\Im\left[\mC_{DuX^2}\right]_{33} S_5+i\frac{s_W^2}{6\pi^2}\frac{m_t^2}{\Lambda^2}\Im\left[\mC_{DuX^2}\right]_{33} S_5,&
    \nn\\
    \left[L_{DuX^2}^L\right]_{pp} = &i\frac{2s_W^2}{27\pi^2}\frac{m_W^2}{\Lambda^2}\Im\left[\mC_{DuX^2}\right]_{33} S_5+i\frac{4s_W^2-3}{24\pi^2}\frac{m_t^2}{\Lambda^2}\Im\left[\mC_{DuX^2}\right]_{33} S_5.&\nonumber
\end{flalign}

\boxtitle{\gamma\text{-}\gamma\text{-}\phi\text{-}\phi}
\begin{flalign}\label{eq:WC:S:gmgm}
    L_{FF\phi^2}=&-\frac{v^2}{\Lambda^2}\frac{2\alpha v}{27\sqrt{2} \pi m_t}\Re\left[\mC_{u\phi^2}\right]_{33},&
    \nn\\
    L_{F \widetilde{F}\phi^2}=&\frac{v^2}{\Lambda^2}\frac{\alpha v}{9\sqrt{2}\pi m_t}\Im\left[\mC_{u\phi^2}\right]_{33}.&
\end{flalign}

\boxtitle{\gamma\text{-}\gamma\text{-}\chi\text{-}\chi}
\begin{flalign}\label{eq:WC:F:gmgm}
    L_{FF\chi}=&-\frac{v^3}{\Lambda^3}\frac{2\alpha v}{27\sqrt{2} \pi m_t}\Re\left[\mC_{u\chi^2}\right]_{33},&
    \nn\\
    L_{F \widetilde{F}\chi}=&\frac{v^3}{\Lambda^3}\frac{\alpha v}{9\sqrt{2} \pi m_t}\Im\left[\mC_{u\chi^2}\right]_{33}.&
\end{flalign}

\boxtitle{\gamma\text{-}\gamma\text{-}X\text{-}X}
\begin{flalign}\label{eq:WC:V:gmgm}
    L_{FFXX}=&-\frac{v^2}{\Lambda^2}\frac{2\alpha v}{27\sqrt{2} \pi m_t}\Re\left[\mC_{uX^2}\right]_{33}-\frac{v^2}{\Lambda^2}\frac{\alpha}{27\pi} \big(\Re\left[\mC_{DqX^2}\right]_{33}
    +\Re\left[\mC_{DuX^2}\right]_{33}\big)S_5,&
    \nn\\
    L_{F \widetilde{F}XX}=&\frac{v^2}{\Lambda^2}\frac{\alpha v}{9\sqrt{2} \pi m_t}\Im\left[\mC_{uX^2}\right]_{33}+i\frac{v^2}{\Lambda^2}\frac{\alpha}{18 \pi }\left(\Re\left[\mC_{DqX^2}\right]_{33}-\Re\left[\mC_{DuX^2}\right]_{33}\right)S_5,&
    \nn\\
    L_{FXFX}=&\frac{v^2}{\Lambda^2}\frac{4\alpha}{27\pi} \big(\Re\left[\mC_{DqX^2}\right]_{33}
    +\Re\left[\mC_{DuX^2}\right]_{33}\big)S_5,&
    \nn\\
    L_{FX\widetilde{F}X}=&i\frac{v^2}{\Lambda^2}\frac{2\alpha}{9 \pi}\big(\Re\left[\mC_{DqX^2}\right]_{33}
    -\Re\left[\mC_{DuX^2}\right]_{33}\big)S_5.&
\end{flalign}

\boxtitle{g\text{-}g\text{-}\phi\text{-}\phi}
\begin{flalign}\label{eq:WC:S:glgl}
    L_{GG\phi^2}=&-\frac{v^2}{\Lambda^2}\frac{\alpha_s v}{12\sqrt{2} \pi m_t}\Re\left[\mC_{u\phi^2}\right]_{33},&
    \nn\\
    L_{G\widetilde{G}\phi^2}=&\frac{v^2}{\Lambda^2}\frac{\alpha_s v}{8\sqrt{2}\pi m_t}\Im\left[\mC_{u\phi^2}\right]_{33}.&
\end{flalign}

\boxtitle{g\text{-}g\text{-}\chi\text{-}\chi}
\begin{flalign}\label{eq:WC:F:glgl}
    L_{GG\chi}=&-\frac{v^3}{\Lambda^3}\frac{\alpha_s v}{12\sqrt{2} \pi m_t}\Re\left[\mC_{u\chi^2}\right]_{33},&
    \nn\\
    L_{G\widetilde{G}\chi}=&\frac{v^3}{\Lambda^3}\frac{\alpha_s v}{8\sqrt{2}\pi m_t}\Im\left[\mC_{u\chi^2}\right]_{33}.&
\end{flalign}

\boxtitle{g\text{-}g\text{-}X\text{-}X}
\begin{flalign}\label{eq:WC:V:glgl}
    L_{GGXX}=&-\frac{v^2}{\Lambda^2}\frac{\alpha_s v}{12\sqrt{2} \pi m_t}\Re\left[\mC_{uX^2}\right]_{33}-\frac{v^2}{\Lambda^2}\frac{\alpha_s}{24\pi} \big(\Re\left[\mC_{DqX^2}\right]_{33}
    +\Re\left[\mC_{DuX^2}\right]_{33}\big)S_5,&
    \nn\\
    L_{G\widetilde{G}XX}=&\frac{v^2}{\Lambda^2}\frac{\alpha_s v}{8\sqrt{2}\pi m_t}\Im\left[\mC_{uX^2}\right]_{33}+i\frac{v^2}{\Lambda^2}\frac{\alpha}{16 \pi }\left(\Re\left[\mC_{DqX^2}\right]_{33}-\Re\left[\mC_{DuX^2}\right]_{33}\right)S_5,&
    \nn\\
    L_{GXGX}=&\frac{v^2}{\Lambda^2}\frac{\alpha_s}{6\pi} \big(\Re\left[\mC_{DqX^2}\right]_{33}
    +\Re\left[\mC_{DuX^2}\right]_{33}\big)S_5,&
    \nn\\
    L_{GX\widetilde{G}X}=&i\frac{v^2}{\Lambda^2}\frac{\alpha_s}{4\pi} \big(\Re\left[\mC_{DqX^2}\right]_{33}
    -\Re\left[\mC_{DuX^2}\right]_{33}\big)S_5.&
\end{flalign}

The loop functions $S_i \equiv S_i(x_t)$ and $V_i \equiv V_i(x_t)$, with $x_t=m_t^2/m_W^2$, involved in the above expressions are defined, respectively, as
\begin{align}
    S_1(x)=&\frac{1}{\tilde\epsilon}-\frac{3(x-3)}{2(x-1)^2}-\frac{3\log x}{(x-1)^3},&
    \nn\\
    S_2(x)=&\frac{2}{\tilde\epsilon}-\frac{x(x-5)-2}{2(x-1)^2}-\frac{3x\log{x}}{(x-1)^3},
    \nn\\
    S_3(x)=&\frac{1}{\tilde\epsilon}-\frac{(x-7)}{2(x-1)}-\frac{3\log{x}}{(x-1)^2},&
    \nn\\
    S_4(x)=&-\frac{1}{\tilde\epsilon} -\frac{1}{3},&
    \nn\\
    S_5(x)=&-\frac{1}{\tilde\epsilon},&\label{eq:loop function:S}
    \\[1em]
    V_1(x)=&-1,&
    \nn\\
    V_2(x)=&\frac{1}{\tilde\epsilon}-\frac{(x+1)^2}{2(x-1)^2}+\frac{(3 x-1)\log{x}}{(x-1)^3},&
    \nn\\
    V_3(x)=&\frac{x(7x-2x^2-11)}{6(x-1)^3}+\frac{x\log{x}}{(x-1)^4},&
    \nn\\
    V_4(x)=&\frac{x(2x-3x^2-19)+8}{6(x-1)^3}+\frac{x (3x-1)\log{x}}{(x-1)^4},&
    \nn\\
    V_5(x)=&\frac{1}{\tilde\epsilon}+\frac{x(2x-7)+11}{6(x-1)^2}-\frac{\log{x}}{(x-1)^3},&
    \nn\\
    V_6(x)=&\frac{2}{\tilde\epsilon}+\frac{x(5-4x)+5}{3(x-1)^2}-\frac{2\log{x}}{(x-1)^3},&
    \nn\\
    V_7(x)=&\frac{2}{\tilde\epsilon}-\frac{x(5x-49)+14}{6(x-1)^2}+\frac{(9x-4)\log{x}}{(x-1)^3},&
    \nn\\
    V_8(x)=&-\frac{1}{\tilde\epsilon}-\frac{x(23-4x)-49}{6(x-1)^2}+\frac{5\log{x}}{(x-1)^3},&
    \nn\\
    V_9(x)=&\frac{x-3x^2}{2(x-1)^2}+\frac{(3x-2)\log x}{(x-1)^3},&
    \nn\\
    V_{10}(x)=&\frac{1}{\tilde\epsilon}+\frac{x-3}{2(x-1)^2}+\frac{\log x}{(x-1)^3},&
\end{align}
where we have combined the divergent and logarithmic parts together and introduce the abbreviation $\frac{1}{\tilde\epsilon}$, which is defined as
\begin{align} 
\frac{1}{\tilde\epsilon} \equiv \frac{1}{\epsilon}-\gamma_E+\log(4\pi) + \log\frac{\mu_{\text{EW}}^2}{m_t^2},
\end{align}
with $\gamma_E$ being the Euler gamma constant. 

\section{Projected flavour constraints}
\label{app:projections}

\begin{figure}[t]
  \centering
  \includegraphics[width=0.96\linewidth]{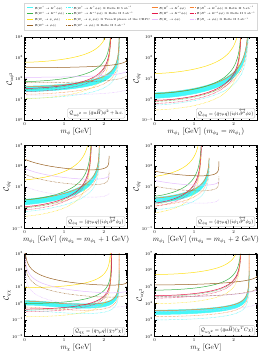}
  \caption{Same as in figure~\ref{fig:flavour:proj}, but for the projected sensitivities to the corresponding operators in figure~\ref{fig:flavour:two}.}
  \label{fig:flavour:two:b2s}
\end{figure}

\begin{figure}[t]
  \centering
  \includegraphics[width=0.96\linewidth]{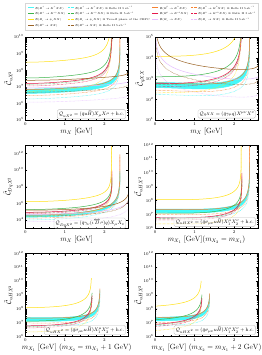}
  \caption{Same as in figure~\ref{fig:flavour:proj}, but for the projected sensitivities to the corresponding operators in figure~\ref{fig:flavour:two:vector}.}
  \label{fig:flavour:two:vector:b2s}
\end{figure}

\begin{figure}[t]
  \centering
  \includegraphics[width=0.96\linewidth]{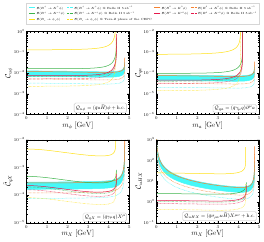}
  \caption{Same as in figure~\ref{fig:flavour:proj}, but for the projected sensitivities to the corresponding operators in figure~\ref{fig:flavour:one}.}
  \label{fig:flavour:one:b2s}
\end{figure}

In this appendix, we collect in figures~\ref{fig:flavour:two:b2s}--\ref{fig:flavour:one:b2s} the projected constraints from the $b\to s$ observables for all DSMEFT operators, complementing the representative examples shown in figure~\ref{fig:flavour:proj}.

\section{Hadronic matrix elements}
\label{app:hadronic matrix element}

The flavour observables in section~\ref{sec:flavour} involve hadronic matrix elements either between a single hadron and the vacuum or between two hadrons, which are parametrized in terms of decay constants and of transition form factors, respectively. Their definitions and conventions are specified in appendices~\ref{sec:decay constant} and~\ref{sec:form factor}. For direct detection in section~\ref{sec:direct detection}, the required quantities are nucleon matrix elements; their definitions and the adopted numerical inputs are provided in appendix~\ref{sec:nucleon form factors}.

\subsection{Decay constants}
\label{sec:decay constant}

For a $\bar B_q$ meson ($q=u,d,s$), only the axial-vector and pseudoscalar currents yield non-vanishing meson-to-vacuum matrix elements, which read
\begin{align}
  \langle 0|\bar{q} \gamma_{\mu} \gamma_5 b | \bar{B}_q (p)\rangle = i f_{B_q} \, p_\mu,
   \quad\quad
  \langle 0|\bar{q} \gamma_{5}  b | \bar{B}_q (p) \rangle = -i \frac{m_{B_q}^2 }{\overline{m}_q(\mu)+\overline{m}_b(\mu)}f_{B_q},
\end{align}
where $f_{B_q}$ denotes the decay constant, while $\overline{m}_b(\mu)$ and $\overline{m}_q(\mu)$ are the $\overline{\text{MS}}$ running quark masses at the scale $\mu$. For a vector quarkonium state $V$ (e.g., $\Upsilon$ or $J/\psi$), the relevant matrix elements are~\cite{Aloni:2017eny,Hatton:2021dvg}
\begin{align}
  \langle 0 | \bar{q} \gamma^\mu q | V(p,\varepsilon) \rangle &= f_V m_V \varepsilon^\mu,\nonumber \\
  \langle 0 | \bar{q} \sigma^{\mu\nu} q | V(p,\varepsilon) \rangle &= if_V^T(\mu) \big ( \varepsilon^\mu p^\nu - \varepsilon^\nu p^\mu\big),
\end{align}
where $\varepsilon^{\mu}$ is the polarization vector of $V$, and $f_V$ and $f_V^T(\mu)$ are its vector and tensor decay constants. Numerical values of all decay constants used in this work, together with their sources, are listed in table~\ref{tab:input}.

\begin{table}[t]
	\center
	\resizebox{\linewidth}{!}{
		\renewcommand\arraystretch{1.3}
		\begin{tabular}{| c | c  c  c | c  c  c | c  c  c |}
            \hline
			\multicolumn{3}{|c|}{$f_{B^0}\,[{\rm GeV}] \quad \qquad  f_{B_s}\,[{\rm GeV}]$~\cite{FlavourLatticeAveragingGroupFLAG:2024oxs}}  & 
			\multicolumn{2}{|c|}{$f_\Upsilon\,[{\rm GeV}]$~\cite{Hatton:2021dvg}} & \multicolumn{2}{|c|}{$f_{J/\Psi}\,[{\rm GeV}]$~\cite{Hatton:2020qhk}} & \multicolumn{3}{|c|}{$f_{\phi_s}\,[{\rm GeV}]~~~~~~~~f^T_{\phi_s}/f_{\phi_s}$~\cite{Konig:2015qat}}
			\\
			\hline
			\multicolumn{3}{|c|}{\quad $0.1905(13)$ \qquad \qquad $0.2307(12)$ ~~~ \qquad}&  
			\multicolumn{2}{|c|}{$0.6772(97) $} & \multicolumn{2}{|c|}{$0.4104(17)$} & \multicolumn{3}{|c|}{\quad $0.2230(14) \qquad \qquad 0.76(4)\qquad $}  
			\\
            \hline
            \multicolumn{2}{|c}{{}$m_t^{\text{pole}}$} & \multicolumn{3}{c}{$(172.52 \pm 0.33)$ GeV~\cite{ParticleDataGroup:2026mpi}}&
            \multicolumn{2}{|c}{$|V_{cb}|$(semi-leptonic) } &
            \multicolumn{3}{c|}{$(41.22 \pm 0.24 \pm 0.37) \times 10^{-3}$~\cite{CKMfitter} }
            \\
            \hline
            \multicolumn{2}{|c}{$|V_{ub}|$(semi-leptonic)} & \multicolumn{3}{c}{$(3.86 \pm 0.07 \pm 0.12) \times 10^{-3} $ GeV~\cite{CKMfitter}}&
            \multicolumn{2}{|c}{$|V_{us}|f^{K \to \pi}_+$(0) } &
            \multicolumn{3}{c|}{$0.2164 \pm 0.0004 $~\cite{CKMfitter} }
            \\
            \hline
            \multicolumn{2}{|c}{$\gamma$ } & \multicolumn{3}{c}{$\big( 66.4^{+2.7}_{-2.8} \big)^\circ $~\cite{CKMfitter}}&
            \multicolumn{2}{|c}{$f^{K \to \pi}_+(0)$} &
            \multicolumn{3}{c|}{$0.9675 \pm 0.0011 \pm 0.0023$~\cite{CKMfitter} }
            \\
			\hline
			\multicolumn{10}{|c|}{$F_1^{q/p}(q^2)$~\cite{Bishara:2017pfq} }
			\\
			\hline
			& $a_{F,\pi}^{u/p}(0)$ & $a_{F,\pi}^{d/p}(0)$ & $a_{F,\pi}^{s/p}(0)$ & $a_{F,\eta}^{u/p}(0)$ & $a_{F,\eta}^{d/p}(0)$ & $a_{F,\eta}^{s/p}(0)$ & $b_{F}^{u/p}(0)$ & $b_{F}^{d/p}(0)$ & $b_{F}^{s/p}(0)$ 
			\\
			\hline
			$F_{P'}^{q/N} (q^2)$  & $2.545(5)$ & $-2.545(5)$ & $0$ & $0.389(26)$ & $0.389(26)$ & $-0.78(5)$  & $-4.65(25)$ & $3.28(25)$ & $0.34(19)$
			\\
			\hline
			$F_{P}^{q/N} (q^2)$  & $3.616(78)$ & $-3.616(78)$ & $0$ & $0.552(39)$ & $0.552(39)$ & $-1.105(78)$ & $1$ & $1$ & $1$
			\\
			\hline
			$F_{\widetilde{G}}^{N} (q^2)$  & $-11.5(4) $ & $-$ & $-$ & $-0.087(6)$ & $-$ & $-$  & $-24.3(1.3)$ & $-$ & $-$
			\\
			\hline
			\multicolumn{7}{|c|}{$F_2^{q/p}(q^2)$~\cite{Bishara:2017pfq} }
			& \multicolumn{3}{c|}{$A_{20}^{q/p}(0)$ ~\cite{Liang:2025kkl,Alexandrou:2020sml,Blumlein:2010rn}}
			\\
			\hline
			& $F_2^{u/p}(0)$ & $F_2^{d/p}(0)$ & $F_2^{s/p}(0)$ & $F_2'^{\,u/p}(0)$ & $F_2'^{\,d/p}(0)$ & $F_2'^{\,s/p}(0)$ &
			$A_{20}^{u/p}(0)$ & $A_{20}^{d/p}(0)$ & $A_{20}^{s/p}(0)$
			\\
			\hline
			$F_1^{q/N}(q^2)$ & $2$ & $1$ & $0$ & $5.57(9)$ & $2.84(5)$ & $-0.018(9)$ &
			$0.359(30)$ & $0.188(19)$ & $0.052(12)$
			\\
			\hline
			$F_2^{q/N}(q^2)$ & $1.609(17)$ & $-2.097(17)$ & $-0.064(17)$ & $-$ & $-$ & $-$ 
			& & &
			\\
			\hline
			$F_A^{q/N}(q^2)$ & $0.897(27)$ & $-0.376(27)$ & $-0.031(5)$ & $-$ & $-$ & $-$ & 
			$\tilde A_{20}^{u/p}(0)$ & $\tilde A_{20}^{d/p}(0)$ & $\tilde A_{20}^{s/p}(0)$
			\\
			\hline
			$F_S^{q/N}(q^2)$ & $7.73$ & $6.81$ & $0.435$ & $-$  & $-$ & $-$ &
			$0.151(4)$ & $-0.039(7)$ & $0.0025(12)$
			\\
			\hline
			$F_G^{N}(q^2)$&$-5.01(19)$&$-$&$-$&$-5.70(21)$&$-$&$-$& &&
			\\
			\hline
		\end{tabular}
	} 
	\caption{Main input parameters used in our numerical analysis. The results for the neutron case can be obtained from those for the proton case by swapping the superscripts $p\leftrightarrow n$ and $u\leftrightarrow d$, under the isospin symmetry. } 
	\label{tab:input}
\end{table}

\subsection{Transition form factors}
\label{sec:form factor}

For FCNC transitions of a $\bar{B}$ meson into a light pseudoscalar meson $P$ (e.g., $\bar{B}\to \bar{K}$), the form factors are defined as~\cite{Beneke:2000wa,Ball:2004ye}
\begin{align}
  c_P \langle P (p^\prime)|\bar{q} \gamma^{\mu} b | \bar{B} (p) \rangle &= f_+(q^2) 
  \left(P^\mu - \frac{m_B^2-m_P^2}{q^2} q^\mu \right) + f_0(q^2) \frac{m_B^2-m_P^2}{q^2}q^\mu, \nonumber\\[0.4em]
  c_P \langle P (p^\prime)|\bar{q} b | \bar{B} (p) \rangle &= \frac{m_B^2-m_P^2}{\overline{m}_b(\mu)-\overline{m}_q(\mu)} f_0(q^2), \nonumber\\[0.4em]
  c_P \langle P (p^\prime)|\bar{q} \sigma^{\mu\nu} b | \bar{B} (p) \rangle &= i \frac{P^\mu q^\nu - P^\nu q^\mu}{m_B+m_P} f_T(q^2),
\end{align}
with $q=p-p^\prime$ and $P=p+p^\prime$; the isospin factor is $c_P=-\sqrt{2}$ for $P=\pi^0$ and $c_P=1$ otherwise. For $\bar{B}\to V$ transitions, with $V$ a light vector meson, the form factors are defined through the Lorentz decompositions~\cite{Beneke:2000wa,Ball:2004rg}
\begin{align}\label{eq:FF:tensor}
  c_V \langle V(p^\prime,\varepsilon^*)|\bar{q} \gamma_{\mu} b | \bar{B} (p) \rangle &= -\epsilon_{\mu\nu\rho\sigma} \varepsilon^{*\nu} P^\rho q^\sigma \frac{V(q^2)}{m_B+m_V}, \nonumber\\[0.4em]
  c_V \langle V(p^\prime,\varepsilon^*)|\bar{q} \gamma^{\mu} \gamma^5 b | \bar{B} (p) \rangle &= i\varepsilon^{*\mu}  (m_B+m_V) A_1(q^2) -iP^\mu \frac{\varepsilon^{*}\cdot q}{m_B+m_V}A_2(q^2) \nonumber\\
  &\qquad -i q^\mu(\varepsilon^*\cdot q)\frac{2 m_V}{q^2}\left[ A_3(q^2)-A_0(q^2) \right],
  \nonumber\\[0.4em]
  c_V \langle V(p^\prime,\varepsilon^*)|\bar{q} \gamma^5 b | \bar{B} (p) \rangle &= -2i\frac{m_V}{\overline{m}_b(\mu)+\overline{m}_q(\mu)} (\varepsilon^* \cdot q) A_0(q^2),
  \nonumber\\[0.4em]
  c_V \langle V(p^\prime,\varepsilon^*)|\bar{q} \sigma_{\mu\nu} q^\nu b | \bar{B} (p) \rangle 
  &= -i\epsilon_{\mu\nu\rho\sigma} \varepsilon^{*\nu} P^\rho q^\sigma T_1 (q^2),
  \nonumber\\[0.4em]
  c_V \langle V(p^\prime,\varepsilon^*)|\bar{q} \sigma_{\mu\nu} q^\nu \gamma^5 b | \bar{B} (p) \rangle 
  &= \left[ (m_B^2-m_V^2) \varepsilon^*_\mu - (\varepsilon^*\cdot q) P_\mu \right] T_2(q^2) \nonumber\\
  &\qquad +  (\varepsilon^*\cdot q) \left (q_\mu - \frac{q^2}{m_B^2-m_V^2} P_\mu\right ) T_3(q^2),
\end{align}
where $\varepsilon$ is the polarization vector of $V$, and $c_V=-\sqrt{2}$ for $V=\rho^0$ and $c_V=1$ otherwise. Throughout, we use $\epsilon^{0123}=+1$ and $\sigma_{\mu\nu}=\tfrac{i}{2}[\gamma_\mu,\gamma_\nu]$. The absence of kinematic singularities at $q^2=0$ in these decompositions requires $f_+(0)=f_0(0)$ and $A_0(0)=A_3(0)$, while the identity $\sigma_{\mu\nu}\gamma_5=\tfrac{i}{2}\epsilon_{\mu\nu\alpha\beta}\sigma^{\alpha\beta}$ implies $T_1(0)=T_2(0)$. Moreover, $A_3$ is not independent and follows from $A_1$ and $A_2$ via~\cite{Ball:2004rg}
\begin{align}
A_3(q^2)=\frac{(m_B+m_V)}{2 m_V} A_1(q^2) - \frac{m_B-m_V}{2 m_V}A_2(q^2).
\end{align}
The uncontracted tensor and axial-tensor matrix elements can be written as~\cite{Khodjamirian:2020btr}
\begin{align}
    c_V\langle V(p^\prime,\varepsilon^*)|\bar{q} \sigma^{\mu\nu} b | \bar{B}(p) \rangle
    =& i\epsilon_{\mu\nu\rho\sigma} \left[ P^\rho\varepsilon^{*\sigma} - \frac{m_B^2 - m_V^2}{q^2} q^\rho \varepsilon^{*\sigma} + \frac{\varepsilon^* \cdot q}{q^2} q^\rho P^\sigma \right] T_1(q^2) \nonumber\\
    & + i\epsilon_{\mu\nu\rho\sigma} \left[\frac{m_B^2 - m_V^2}{q^2} q^\rho \varepsilon^{*\sigma} + \frac{\varepsilon^* \cdot q}{q^2} P^\rho q^{\sigma} \right] T_2(q^2) \nonumber\\
    & + i\epsilon_{\mu\nu\rho\sigma} \frac{\varepsilon^* \cdot q}{m_B^2 - m_V^2} P^\rho q^{\sigma} T_3(q^2) ,
    \nonumber\\[0.2cm]
    c_V\langle V(p^\prime,\varepsilon^*)|\bar{q} \sigma^{\mu\nu} \gamma^5 b | \bar{B}(p) \rangle
    =& \bigg[(\varepsilon^{*\mu} P^\nu - P^\mu \varepsilon^{*\nu}) 
    + \frac{m_B^2 - m_V^2}{q^2} (q^\mu \varepsilon^{*\nu} - \varepsilon^{*\mu} q^\nu) \nonumber\\
    & + \frac{\varepsilon^*\cdot q}{q^2} (P^\mu q^{\nu} - q^{\mu} P^\nu)\bigg]T_1(q^2) \nonumber\\
    &\hspace{-1.3cm} + \left[\frac{m_B^2 - m_V^2}{q^2} (\varepsilon^{*\mu} q^\nu - q^\mu \varepsilon^{*\nu}) + \frac{\varepsilon^*\cdot q}{q^2} (q^{\mu} P^\nu - P^\mu q^{\nu})\right]T_2(q^2) \nonumber\\
    & + \frac{\varepsilon^*\cdot q}{m_B^2 - m_V^2} \left(q^{\mu} P^\nu - P^\mu q^{\nu}\right) T_3(q^2),
\end{align}
which reduce to the last two relations of eq.~\eqref{eq:FF:tensor} upon contraction with $q_{\nu}$. Form factors for $B_s \to P\,(V)$ and $K\to\pi$ transitions are defined analogously.

Since lattice QCD results are not yet available for the complete set of these form factors, notably those of the tensor currents (see the FLAG review~\cite{FlavourLatticeAveragingGroupFLAG:2024oxs} and references therein), we adopt the light-cone sum rule predictions of ref.~\cite{Ball:2004ye} for $B \to P$ and of ref.~\cite{Ball:2004rg} for $B \to V$ transitions. For the $K\to \pi$ transition, the vector and scalar form factors are taken from refs.~\cite{Bernard:2006gy,Bernard:2009zm}, and the tensor one from ref.~\cite{Baum:2011rm}.

No calculation of the form factors for the operators $\mO_{DdX^2}^{L,R}$ in table~\ref{tab:DLEFT} is available in the literature. Assuming the Wilson coefficients $L_{DdX^2}^{L,R}$ to be hermitian in flavour space, however, these form factors can be expressed through those given above. For $\bar B\to\bar K^0$, for instance, one has
\begin{align}
  \langle \bar K^0(p^\prime) \lvert (\partial^\mu \bar s) \gamma^\nu P_L b + \bar s \gamma^\nu P_L (\partial^\mu b) \rvert \bar B(p) \rangle = - i q^\mu \langle \bar K^0(p^\prime) \lvert  \bar s \gamma^\nu P_L b \rvert \bar B(p) \rangle.
\end{align}

\subsection{Nucleon form factors}
\label{sec:nucleon form factors}

We provide the parametrizations of the quark bilinear operators used in our numerical analysis of direct detection. The general parametrizations of quark-current matrix elements between nucleon states read~\cite{DelNobile:2021wmp,Adler:1975he,Ji:1998pc}:
\begin{align} \label{eq:hadff}
			\langle N(k')| G^{a}_{\mu\nu} G^{\mu\nu,a} |N(k) \rangle & =F_G^N \,\bar u_{N'}u_N 
            \nn\\
			\langle N(k')| G^{a}_{\mu\nu} \widetilde G^{\mu\nu,a} |N(k) \rangle & =F_{\widetilde G}^N \,\bar u_{N'}i \gamma_5 u_N \nn
			\\
			\langle N(k')| \bar q q | N(k)\rangle & = 
			F_S^{q/N}\, \bar u_{N'}u_N, \nn
			\\
			\langle N(k')| \bar q i\gamma_5 q | N(k)\rangle & =
			F_P^{q/N}\, \bar u_{N'}i \gamma_5 u_N, \nn
			\\
			\langle N(k')| \bar q \gamma_\mu q | N(k)\rangle & =
			\overline{u}_{N'} \left[
			F_{1}^{q/N}\, \gamma_\mu  + F_{2}^{q/N}\,  \frac{i\sigma_{\mu\nu} q^\nu}{2m_{N}}  
			\right] u_{N} , \nn
			\\
			\langle N(k') |  \bar q \gamma_{\mu}\gamma_5 q | N(k) \rangle
			&=
			\overline{u}_{N'} \bigg[ 
			F_A^{q/N}\, \gamma_\mu\gamma_5 + 
			F_{P'}^{q/N}\, \frac{ q_\mu}{2m_{N}} \gamma_5
			\bigg] u_{N}, \nn
			\\
			\langle N(k')| ( \overline{q}\gamma_{(\mu} i \overleftrightarrow{D}_{\nu)} q) | N(k) \rangle
			& = \overline{u}_{N'} \left[  
			A_{20}^{q/N}\, \gamma_{(\mu} K_{\nu)} + B_{20}^{q/N}\, \frac{i \sigma_{(\mu\rho} K_{\nu)} q^{\rho}}{2 m_N} \right. \nn \\
            & \left. \hspace{1.5cm} + C_{20}^{q/N}\, \frac{ 2 q_{(\mu}q_{\nu)}}{m_N}\right] u_{N}, \nn
			\\
			\langle N(k')| ( \overline{q}\gamma_{(\mu}\gamma_5 i \overleftrightarrow{D}_{\nu)} q) | N(k) \rangle
			& = \overline{u}_{N'} \left[  
			\tilde A_{20}^{q/N}\, \gamma_{(\mu}\gamma_5 K_{\nu)}  + \tilde B_{20}^{q/N}\, \frac{K_{(\mu}q_{\nu)}}{2 m_N}\gamma_5  \right] u_{N},
\end{align}
with $q^\mu\equiv k'^\mu-k^\mu$, and $K^\mu\equiv k'^\mu+k^\mu$; $u_{N}\equiv u_N(k)$ and $\bar{u}_{N'} \equiv \bar{u}_N(k')$ are the Dirac spinors of the incoming and outgoing nucleons, and the form factors depend on $q^2$ throughout $I^{q/N}_i\equiv I^{q/N}_i(q^2)$. The traceless symmetrization entering the last two currents is defined as $ A_{(\mu } B_{\nu)} = \frac{1}{2}(A_{\mu}B_{\nu} + A_{\nu}B_{\mu}) - \frac{1}{4}g_{\mu\nu} A_\rho B^\rho$.

The leading-order (LO) non-relativistic reduction of the above matrix elements has been performed in ref.~\cite{Liang:2025kkl}, with the resulting correspondence:
\begin{align} \label{eq:LO:NR}
		\bar q q & \Rightarrow 2 m_N F_S^{q/N} \mathds{1}_N, \nn
		\\[0.15cm]
		\bar q i\gamma_5 q  &  \Rightarrow  - 2 F_P^{q/N} i \pmb{q}\cdot \pmb{S}_N, \nn
		\\[0.15cm]
		\bar q \gamma_\mu q & \Rightarrow  F_{1}^{q/N} \left( 2 m_N g_\mu^0 \mathds{1}_N 
		-  g_\mu^i K^i \right)
		+ 2 \left(F_{1}^{q/N} +F_{2}^{q/N} \right) g_\mu^i (i \pmb{q}\times\pmb{S}_N)^i, \nn
		\\[0.15cm]
		\bar q \gamma_{\mu}\gamma_5 q & \Rightarrow  
		- 4 m_N F_A^{q/N} g_\mu^i S_N^i 
		+ 2 F_A^{q/N} g_\mu^0 \pmb{K}\cdot \pmb{S}_N
		- 2 F_{P'}^{q/N} \frac{q_\mu}{m_N} \pmb{q}\cdot \pmb{S}_N, \nn
		\\[0.15cm]
		\overline{q}\gamma_{(\mu} i \overleftrightarrow{D}_{\nu)} q
		&\Rightarrow 
		4 m_N^2 A_{20}^{q/N}\left(g_\mu^0 g_\nu^0 - \frac{1}{4} g_{\mu\nu}\right)  \mathds{1}_N, \nn
		\\[0.15cm]
		\overline{q}\gamma_{(\mu}\gamma_5 i \overleftrightarrow{D}_{\nu)} q
		&\Rightarrow 
		2 m_N \tilde A_{20}^{q/N} \left[ - 2 m_N (g_\mu^0 g_\nu^i + g_\mu^i g_\nu^0 )
		+(2 g_\mu^0 g_\nu^0 \delta^{ij} + g_\mu^i g_\nu^j+g_\mu^j g_\nu^i) K^j \right] S_N^i.
\end{align}

Most of the nucleon form factors entering eq.~\eqref{eq:hadff} have been determined within chiral perturbation theory (ChPT)~\cite{Bishara:2017pfq,Liang:2024tef}. In this framework, they account for two classes of contributions to DM--nucleon scattering: local DM--nucleon interactions and $t$-channel exchanges of intermediate mesonic degrees of freedom. The form factors $F_P$ and $F_{P'}$ receive their LO contributions from $\pi$- and $\eta$-exchange diagrams, while the contact DM--nucleon interactions enter only at next-to-leading (NLO) in the chiral power counting. Denoting these two form factors collectively as $\mathcal{F}_1^{q/N}(q^2)$, we write them as a sum of two pole terms and a contact term,
\begin{align}
	\mathcal{F}_1^{q/N}(q^2) &= \frac{m_N^2}{m_\pi^2-q^2} a_{{\mathcal{F}_1},\pi}^{q/N} + \frac{m_N^2}{m_\eta^2-q^2} a_{{\mathcal{F}_1},\eta}^{q/N} + b_{\mathcal{F}_1}^{q/N},\quad\text{for}~ \mathcal{F}_1= F_{P'}, F_P, F_{\widetilde{G}}^N.
\end{align}
The remaining form factors, $F_{1,2}^{q/N}$, $F_A^{q/N}$, $F_S^{q/N}$, and $F_G^N$, parametrize the contributions of the local DM--nucleon interactions. Collecting them as $\mathcal{F}_2^{q/N}(q^2)$, their chiral expansion up to NLO can be written as
\begin{align}
	\mathcal{F}_2^{q/N}(q^2) &= \mathcal{F}_2^{q/N}(0) +\mathcal{F}_2'^{q/N}(0) \frac{q^2}{{\rm GeV}^2}.
\end{align}
The numerical values of the parameters entering $\mathcal{F}_1^{q/p}(q^2)$ and $\mathcal{F}_2^{q/p}(q^2)$ for the proton are listed in table~\ref{tab:input}, following the summary of ref.~\cite{Bishara:2017pfq}. The corresponding values for the neutron follow from isospin symmetry by exchanging the superscripts $p\leftrightarrow n$ and $u\leftrightarrow d$. For the two derivative bilinears $\overline{q}\gamma_{(\mu} i \overleftrightarrow{D}_{\nu)} q$ and $\overline{q}\gamma_{(\mu}\gamma_5 i \overleftrightarrow{D}_{\nu)} q$ in eq.~\eqref{eq:LO:NR}, only the form factors $A_{20}^{q/N}$ and $\tilde A_{20}^{q/N}$ enter our calculation, for which we adopt the input values from refs.~\cite{Liang:2025kkl,Alexandrou:2020sml,Blumlein:2010rn}.

\bibliographystyle{JHEP}

\begingroup
\setlength{\bibsep}{5pt}
\bibliography{ref}
\endgroup

\end{document}